\documentclass[12pt,psf,epsf]{article}
\usepackage[centertags]{amsmath}
\usepackage{amssymb}
\usepackage{mathtools}
\usepackage{graphicx,dsfont}
\usepackage{epsfig}
\usepackage{ulem}
\usepackage[english]{babel}
\usepackage{array}
\usepackage{amsthm}
\usepackage{latexsym}
\usepackage[mathcal]{euscript}
\pdfoutput=1
\usepackage{epsfig}
 \usepackage{jheppubm}
\usepackage{mathdots}
\usepackage{MnSymbol}
\usepackage{multirow}
\usepackage{comment}
\usepackage[export]{adjustbox}

\newcommand{\be}{\begin{equation}}
\newcommand{\ee}{\end{equation}}
\newcommand{\bea}{\begin{eqnarray}}
\newcommand{\eea}{\end{eqnarray}}

\newcommand{\e}{\mathrm{e}}

\newcommand{\mI}{\mathcal{I}}

\newcommand{\mO}{\mathcal{O}}

\newcommand{\mH}{\mathcal{H}}

\newcommand{\bZ}{{\bf Z}}

\DeclareMathOperator{\tr}{tr}
\DeclareMathOperator{\Tr}{Tr}

\definecolor{dgreen}{rgb}{0.,0.6,0.}

\title{Factorization of the Hilbert space of eternal black holes in general relativity}

\author[1,2,3,4]{Vijay Balasubramanian,}
\author[2]{Ben Craps,}
\author[2]{Juan Hernandez,}
\author[1,2]{Mikhail Khramtsov,}
\author[2]{Maria Knysh}

\abstract{
We generalize recent results in two-dimensional Jackiw-Teitelboim gravity  to study factorization of the Hilbert space of eternal black holes in quantum gravity with a negative cosmological constant in any dimension. We approach the problem by computing the  trace  of two-sided observables as a sum over a recently constructed family of semiclassically well-controlled black hole microstates. These microstates, which contain heavy matter shells behind the horizon and form an overcomplete basis of the  Hilbert space, exist in any theory of gravity with general relativity as its low energy limit. Using this representation of the microstates, we show that the  trace of operators dual to functions of the Hamiltonians of the  left and right holographic CFTs factorizes into a product over left and right factors to leading order in the semiclassical limit.  Under certain conditions this implies factorization of the Hilbert space.
}

\affiliation[1]{David Rittenhouse Laboratory, University of Pennsylvania, 209 S. 33rd Street, Philadelphia, PA 19104, USA}
\affiliation[2]{Theoretische Natuurkunde, Vrije Universiteit Brussel (VUB) and The International Solvay Institutes, Pleinlaan 2, B-1050 Brussels, Belgium}
\affiliation[3]{Santa Fe Institute, 1399 Hyde Park Road, Santa Fe, NM 87501, USA}
\affiliation[4]{Rudolf Peierls Centre for Theoretical Physics, University of Oxford, Oxford OX1 3PU, United Kingdom}

\emailAdd{vijay@physics.upenn.edu}
\emailAdd{ben.craps@vub.be}
\emailAdd{juan.hernandez@vub.be}
\emailAdd{mikhail.khramtsov@vub.be}
\emailAdd{maria.knysh@vub.be} 

\date{June 2024}

\begin{document}

 \maketitle
\section{Introduction}
In the AdS/CFT correspondence, a two-sided black hole, or Einstein-Rosen bridge, in asymptotically AdS spacetime is dual to a state defined in a double copy of the holographic CFT~\cite{Balasubramanian:1998de,Maldacena:2001kr}.  
The thermofield double (TFD) state is an example of this correspondence, where the entanglement between the two CFTs is understood to be geometrized in the dual description by the wormhole behind the horizon connecting two asymptotic regions \cite{VanRaamsdonk:2010pw}. The effective field theory (EFT) of fluctuations on this connected geometry does not have a Hilbert space that factorizes into a product of components, unlike the underlying CFT. This is not surprising since subspaces of a factorized Hilbert space can of course be entangled.  However, the full Hilbert space contains more general geometries, connected and disconnected, together with the corresponding Hilbert spaces of quantum fluctuations. The question of how this bigger space separates into a product of factors is referred to as the Hilbert space factorization puzzle.

This factorization puzzle has recently been addressed in~\cite{Boruch:2024kvv} for holographic systems dual to JT (Jackiw–Teitelboim) gravity in asymptotically AdS$_2$ spacetime, using results derived in~\cite{Iliesiu:2022kny,Boruch:2023trc,Iliesiu:2024cnh}. Specifically, the authors of \cite{Boruch:2024kvv} give evidence that the bulk Hilbert space $\mH_{\text{bulk}}$ of the two-sided JT black hole factorizes into a tensor product $\mH_L \otimes \mH_R$, where $\mH_{L,\,R}$ correspond to  Hilbert spaces of holographically dual theories living on the left and right boundaries. To this end, they consider the trace of a product of bulk duals of operators $k_L$ and $k_R$ drawn from the algebras of observables acting on $\mH_{L,\,R}$, respectively. Their main result is the factorization of this trace\footnote{Note that \cite{Boruch:2024kvv} uses two different spellings of the word ``factorization" to distinguish the factorization of the Hilbert space from the factorization of Euclidean gravity observables on a set of disconnected asymptotic boundaries. In the present work, we are mostly focused on the former and we will provide explicit context when commenting on the latter, so we will not follow the same naming convention.} over the bulk Hilbert space to all (perturbative and nonperturbative) orders in the Newton constant $G_N$,
\be
\Tr_{\mH_{\text{bulk}}} k_L k_R = \Tr_{\mH_{L}} k_L \Tr_{\mH_{R}} k_R\,.\label{eq: factorization-general}
\ee
In this equation we have employed the holographic dictionary which maps the boundary operators $k_L$, $k_R$ on the right hand side to their bulk dual images on the left hand side. We denote the bulk images of operators with the same symbols as their respective boundary operators. In \cite{Boruch:2024kvv} the equality (\ref{eq: factorization-general}) is verified by means of a path integral for which JT gravity provides an effective field theory (EFT) capturing the low energy degrees of freedom. This approach to gravity, where the EFT path integral is treated as a coarse-grained approximation to a UV-complete theory of quantum gravity, has been instrumental in investigating various questions of quantum gravity, including the black hole information problem \cite{Penington:2019kki,Almheiri:2019qdq}. The article \cite{Boruch:2024kvv} shows that, surprisingly, in JT gravity this coarse-grained description is enough to show factorization of the bulk Hilbert space into a product of left and right Hilbert spaces.

Here, we extend the analysis of \cite{Boruch:2024kvv}  to holographic systems whose low energy degrees of freedom are captured by General Relativity (GR) in asymptotically AdS$_{d+1}$ for any $d\geq 2$. This effective low energy description can be thought of as descending from a UV completion, be it string theory or some other theory of quantum gravity. We adopt the techniques and framework developed in~\cite{Balasubramanian:2022gmo,Balasubramanian:2022lnw,Climent:2024trz}, and show  that the overall logic of~\cite{Boruch:2024kvv} carries through in higher dimensions.  

The key ingredient is the enumeration of an explicit basis of black hole microstates.  Recent work has shown how this Hilbert space of microstates can be  spanned by a basis whose elements have semiclassical descriptions in terms of shells of matter propagating behind the horizon \cite{Balasubramanian:2022gmo,Balasubramanian:2022lnw,Climent:2024trz}.  These microstates, which have small quantum overlaps in analogy to the coherent state basis of the harmonic oscillator, exist in any UV-complete theory with GR as its low-energy limit, in any dimension, and for black holes of any mass, charge and angular momentum.  As such, they generalize constructions in 2d gravity \cite{Penington:2019kki,Almheiri:2019qdq,Saad:2019pqd,Goel2019,Iliesiu:2022kny,Boruch:2023trc,Iliesiu:2024cnh} as well as related results~\cite{Sasieta:2022ksu,Chandra:2022fwi,Chandra:2022bqq,Dong:2024tjx}.  For universes that are described by string theory, the strings and branes \cite{Strominger:1996sh} or fuzzballs \cite{Bena:2022rna} that appear in the usual descriptions of extremal, supersymmetric black hole microstates should be understood as superpositions in the basis of \cite{Balasubramanian:2022gmo,Balasubramanian:2022lnw,Climent:2024trz}. Indeed, macroscopically different gravitational states can be superposed in this way to give other states with semiclassical descriptions -- simple examples and criteria appear in \cite{Balasubramanian:2007zt}. In the gravitational description every element of the microstate basis in \cite{Balasubramanian:2022gmo,Balasubramanian:2022lnw,Climent:2024trz} has a geometric wormhole behind the horizon connecting two asymptotic regions. Thus the factorization puzzle \cite{Harlow2016a,Harlow2018,Banerjee2022}\footnote{This should not be confused with related factorization puzzle of Euclidean gravity observables \cite{Maldacena:2004rf,Saad212,Saad2021,Cotler2021,Blommaert2021, Belin2021,Penington:2023dql,Hernandez-Cuenca2024} on a set of disconnected spacetime boundaries.  Yet another kind of factorization in gravity is the approximate one associated to the interior and exterior of a black hole.  This has been argued to arise because the complexity of the operators creating black hole microstates and effective field theory states behind the horizon  protects them to exponential accuracy from  ``errors'' created by the action of the simple operators that act on the exterior Hilbert space, see, e.g.,  \cite{Balasubramanian:2018yjq,Balasubramanian:2022fiy}.} naively persists.  We will follow the methods of \cite{Boruch:2024kvv} and use the microstates of \cite{Balasubramanian:2022gmo,Balasubramanian:2022lnw,Climent:2024trz} to demonstrate explicit factorization of traces of observables, which under certain conditions implies a factorized Hilbert space.

There are two main new subtleties relative to the logic in \cite{Boruch:2024kvv}. The first one concerns the structure of  nonperturbative corrections to the gravity path integral. In JT gravity, all of these are well understood \cite{Saad:2019lba}, but this is not generally the case in higher-dimensional gravity. Therefore, we consider only the first non-trivial nonperturbative corrections in the semiclassical expansion. The second subtlety is that we have to consider microcanonical energy windows in our analysis. In JT gravity the exact spectral density and overlap statistics are known \cite{Goel2019}, so moments of the state overlaps can be computed and microstate counting can be performed for energy bands of arbitrary size. While the overlap statistics in higher-dimensional GR can be computed in general,
the microstate counting procedure outlined in \cite{Balasubramanian:2022gmo,Balasubramanian:2022lnw,Climent:2024trz}
is only tractable in the microcanonical ensemble. Thus we have to include an extra step of extending our argument from the microcanonical window to energy windows of arbitrary size.

Three sections follow. In Sec.~\ref{sec:microcanonical} we introduce our analysis framework and provide evidence for the factorization of the microcanonical Hilbert space.  Sec.~\ref{sec: recap} outlines the argument for factorization in~\cite{Boruch:2024kvv} which will be the backbone of our approach. Sec.~\ref{Basis of states} introduces the semiclassically well-defined black hole microstates we use to show factorization of the trace of products of functions of  bulk duals of the left and right Hamiltonians in Secs.~\ref{Trace factorization} and \ref{Trace differential}.  Section~\ref{sec:proof} explains why trace factorization implies Hilbert space factorization and extends the result to arbitrary sized energy bands. The paper concludes with a discussion in Sec.~\ref{sec: discuss}.
Appendix~\ref{sec:LinearAlgebra} recalls how to use overcomplete bases to compute traces.  Details about the black hole microstates used in our arguments appear in Appendix~\ref{app:states and gram}, while Appendix~\ref{tiny calculation} contains additional details of the overlap computations.
Appendix~\ref{sec: generic} discusses extensions of our trace factorization analysis to general light operators.

\section{Factorization of the microcanonical Hilbert space}\label{sec:microcanonical}

\subsection{Outline of the argument}\label{sec: recap}

First, we outline the idea of the calculation following  \cite{Boruch:2024kvv}. We start by considering operators  $k_L(H_L)$ and $k_R(H_R)$ where $H_{L, \,R}$ are the bulk AdS duals to Hamiltonians of the left and right copies of the dual boundary CFT.  For example, we can pick
\be
k_{L} = e^{-\beta_{L} H_{L}}\,,\qquad k_{R} = e^{-\beta_{R} H_{R}}\,,
\label{eq:ExpHamOps}
\ee 
where $\beta_{L,\,R}$ are arbitrary positive parameters. The goal is to show that 
\be
Z_{\text{bulk}} = \Tr_{\mH_{\text{bulk}}} k_L k_R = \Tr_{\mH_L}  k_L\Tr_{\mH_R} k_R\,. \label{eq: factorization-main}
\ee
Here, $\mH_{\text{bulk}}$ is the bulk Hilbert space, while
$\mH_{L,\,R}$  are two Hilbert space factors. The two-sided partition function $Z_{\rm bulk}$ is a generalization of the usual partition function ${\rm Tr}\left( e^{-\beta H_{\rm bulk}}\right)$. The bulk Hamiltonian $H_{\rm bulk}=H_L+H_R$ is a sum of boundary contributions and in our generalization we are allowing for different left and right temperatures.

The first step is to decompose the boundary Hilbert space into microcanonical windows $\mH_{L,\,R} = \bigoplus_E \mH^E_{L,\,R}$, where $E$ labels the microcanonical band $[E_{L,\, R},\, E_{L, \,R} + \delta E)$. The width $\delta E$  of the band is assumed to be small compared to the temperature scale $\beta^{-1}_{L\,,R}$, but large compared to the average energy spacing $\epsilon$. Thus each microcanonical band in the boundary Hilbert space has a width limited by
\begin{equation}
     \epsilon \ll \delta E\ll \beta^{-1}_{L\,,R}\,. \label{eq: limits on width}
\end{equation}
The holographic dictionary guarantees that the partitioning of the boundary Hilbert space into microcanonical bands carries over to the bulk Hilbert space~\cite{Balasubramanian:1998sn,Witten:1998qj}. Thus the bulk two-sided partition function can be written as a sum of microcanonical partition functions of each energy band. In this case, we can write
\be
Z_{\text{bulk}} = \sum_{E_L,\, E_R}\ \bZ(E_L, E_R; \beta_L, \beta_R)\,,
\ee
where we define the microcanonical bulk trace 
\be
\bZ(E_L, E_R; \beta_L, \beta_R) = e^{-\beta_L E_L} e^{-\beta_R E_R} \Tr_{\mH_{\text{bulk}}^E} \left(\mathds{1}\right) = e^{-\beta_L E_L} e^{-\beta_R E_R}\dim(\mH_{\text{bulk}}^E)\,.\label{eq: bZ-def}
\ee
We will sometimes use the shorthand notation $\bZ(E; \beta)$, where $E\equiv (E_L, E_R)$ and $\beta\equiv(\beta_L, \beta_R)$. As is clear from \eqref{eq: bZ-def}, $\bZ(E; \beta)$ factorizes if and only if the dimension of the microcanonical bulk Hilbert space factorizes. The dimension of this Hilbert space has been studied recently, e.g., in \cite{Balasubramanian:2022gmo, Balasubramanian:2022lnw}, using the gravitational path integral for the case where $E_L = E_R$. In what follows, we extend these results to the microcanonical window with fixed but arbitrary left and right energies. However, we will follow the approach in \cite{Boruch:2024kvv}. Along the way, we will introduce the necessary techniques to show that the microcanonical two-sided partition function factorizes into a product of left and right components. We proceed as follows to leading order in $e^{-1/G_N}$: 
\begin{itemize}
    \item[1.]  In Sec.~\ref{Trace factorization} we follow the logic in \cite{Boruch:2024kvv} to show that $\overline{\bZ(E; \beta)}$ factorizes between left and right parts. The overbar here indicates a computation using the semiclassical saddle points of the effective low energy path integral, which we understand as averaging over the UV degrees of freedom. In other words, this section will give a coarse-grained argument for factorization of the quantum gravity Hilbert space.
    \item[2.] To extend the factorization argument to the fine-grained level, we generalize \cite{Boruch:2024kvv} in Sec.~\ref{Trace differential} by using the differential matrix
\begin{equation}\label{eq: diff-intro}
    d(E; \beta)_{\alpha_L\,\alpha_R} = \bZ (E; \beta) \partial_{\alpha_L} \partial_{\alpha_R} \bZ (E; \beta)-\partial_{\alpha_L} \bZ (E; \beta) \partial_{\alpha_R} \bZ(E; \beta) \,,
\end{equation}
where $\alpha_{L,R} \in \{\beta_{L,R},E_{L,R}\}$. This differential is a two-by-two matrix that vanishes if and only if the total trace factorizes between left and right parts\footnote{\label{foot:dE=0}This can be proved as follows. We begin by defining $U=\ln \bold{Z}$. Then the equation $d(E;\beta)_{\alpha_L\,\alpha_R}=0$ can be rewritten as  
\begin{equation}
     \partial_{\alpha_L} \partial_{\alpha_R} U =0\,.
\end{equation}
This is solved by $U= a(E_L,\beta_L)+b(E_R,\beta_R)$ with $a(E_L,\beta_L)$ and $b(E_R,\beta_R)$ some arbitrary functions. Therefore $d=0$ if and only if $\bold{Z}=e^{a(E_L,\beta_L)+b(E_L,\beta_R)}$ is a factorized product of arbitrary functions of left and right energies and temperatures as in \eqref{factor}.}
\begin{equation}\label{factor}
     d(E;\beta)=0 \Longleftrightarrow \bZ(E;\beta) = g_R(E_L; \beta_L) g_L(E_R; \beta_R) 
     \,.
\end{equation}
Furthermore, considering the cases $\beta_L=0$ and $\beta_R=0$, one deduces that the functions $g_{L,R}$ are proportional to the bulk trace over $k_{L,R}$, respectively, such that
\begin{equation}\label{eq: important}
     d(E;\beta)=0 \Longleftrightarrow \bZ(E;\beta) = \frac{{{\rm Tr}_{{\cal H}^{E}_{\rm bulk}}\left( k_L\right)} {{\rm Tr}_{{\cal H}^{E}_{\rm bulk}}\left( k_R\right)} }{ {{\rm Tr}_{{\cal H}^E_{\rm bulk}}\left( \mathds{1}\right)}} 
     \,.
\end{equation}
As compared to \cite{Boruch:2024kvv}, where a simple differential $d(\beta)$ was used, we have to use a  differential matrix because of the dependence of $\bZ(E;\beta)$ on the microcanonical energies $E$. We show that $\overline{\left(d(E;\beta)_{\alpha_L\,\alpha_R}\right)^2 }= 0$, where the overbar again indicates computations using the semiclassical saddle points of the gravity path integral. The vanishing of the second moment of each component of the differential matrix establishes that $d(E; \beta)$ is identically zero, at least to the order that our approximation is valid.  This in turn implies that the microcanonical two-sided partition function of the low energy degrees of freedom factorizes into a product of identical functions that depend on left and right quantities as given in  \eqref{eq: important}. Sec.~\ref{sec:proof} shows that this is a sufficient condition to show the factorization of the bulk Hilbert space under certain assumptions.

The crucial ingredient we will exploit to show the vanishing of the differential is the factorization of the coarse-grained product of microcanonical two-sided partition functions  to leading order in $e^{-1/G_N}$
\begin{equation}
    \overline{ \bZ(E^1;\beta^1)\ldots \bZ(E^n; \beta^n)}=\overline{\bZ(E^1;\beta^1)} \ldots \overline{\bZ(E^n;\beta^n)}  
    \,.
\end{equation}
This implies that fluctuations of the partition function around its coarse-grained value vanish
\begin{equation}
    \overline{\left(\bZ(E;\beta) - \overline{\bZ(E;\beta)}\right)^2} =0
\end{equation}
to leading order in $e^{-1/G_N}$. In turn, this implies that the microcanonical partition function is equal to its coarse-grained value
\begin{equation}
    \bZ(E;\beta)=\overline{\bZ(E;\beta)}\,.
\end{equation}
Thus the microcanonical two-sided partition function of the low energy degrees of freedom factorizes into a product of traces of left and right operators, computed semiclassically to leading order in $e^{-1/G_N}$.

    \item[3.] In Sec. \ref{sec:proof}, we explain how the factorization of the trace implies the Hilbert space factorization. As a byproduct of the discussion, we also explain the generalization of the factorization from a microcanonical window to an energy window of arbitrary finite size.  
\end{itemize}

\subsection{Basis of states}\label{Basis of states}
The key step in the procedure outlined above is summarized by equation \eqref{eq: bZ-def}, where we must compute the dimension of the bulk Hilbert space in a microcanonical band.  To do so, we follow \cite{Balasubramanian:2022gmo,Balasubramanian:2022lnw} and consider certain semiclassically well-defined bulk states dual to CFT states on the tensor product of two copies of the CFT$_d$ with the same Hamiltonian $H$, so that $H_L =H_R=H$. Each copy of the CFT lives on a spacetime of topology $\mathbb{S}^{d-1}\times \mathbb{R}$.  The corresponding bulk states are double sided black holes in asymptotically AdS$_{d+1}$ spacetime with shells of matter behind the horizon.  These states span a bulk Hilbert space ${\cal H}_{\rm bulk}$ which is dual to the Hilbert space of the double copy of the CFT $\mathcal{H}^{CFT}_L \otimes \mathcal{H}^{CFT}_R$. Since each of the semiclassical bulk states has a wormhole connecting the two asymptotic boundaries, we still have a puzzle: how does the bulk Hilbert space factorize?
\begin{figure}[t]
    \centering
    \includegraphics[scale=0.5]{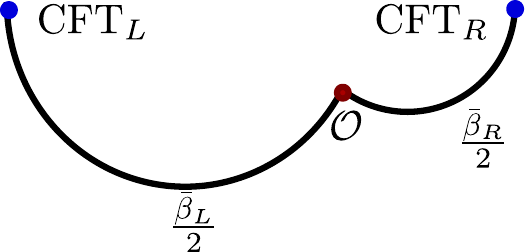} \quad 
    \includegraphics[scale=0.5]{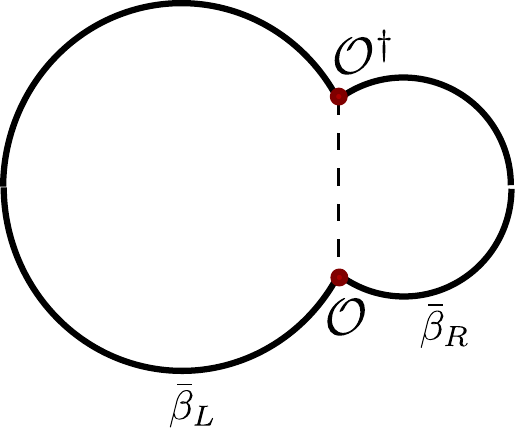} \\ 
    (a) \hspace{5cm} (b)
    \caption{(a) Path integral contour that prepares the state given in eq.~(\ref{state}). The lengths of the circular segments are given by the Euclidean preparation times $\Bar{\beta}_L/2$ and $\Bar{\beta}_R/2$. (b) Euclidean time contour which computes the norm squared of the state given in eq.~(\ref{state}). This quantity can be computed using the gravitational path integral evaluated for geometries with asymptotic boundary length ${\Bar{\beta}_L + \Bar{\beta}_R}$ and with matter excitations sourced by the operators $\mO$ and $\mO^\dagger$. In both figures there is a transverse sphere $\mathbb{S}^{d-1}$ at every point.} 
    \label{fig:contour}
\end{figure}

The energy basis of the doubled CFT Hilbert space $\mathcal{H}^{CFT}_L \otimes \mathcal{H}^{CFT}_R$ is 
\begin{equation}
    (H_L \otimes \mathds{1}) |m \rangle_L \otimes |n \rangle_R=E_m|m \rangle_L \otimes |n \rangle_R\,,\qquad (\mathds{1} \otimes H_R)|m \rangle_L \otimes |n \rangle_R=E_n|m \rangle_L \otimes |n \rangle_R\,.
\end{equation}
We can consider states in this Hilbert space of the form 
\be
| \Psi \rangle = \frac{1}{\sqrt{Z_1}} \sum_{n, m} e^{-\frac12 \Bar{\beta}_L E_m - \frac12 \Bar{\beta}_R E_n} \mO_{mn} |m \rangle_L \otimes |n \rangle_R\,, \label{state}
\ee
where $\mO$ is a primary operator in a single copy of the CFT, $\mO_{mn} = \langle m | \mO | n \rangle$ and $Z_1=\text{Tr}\left[ \mO^\dagger e^{- \Bar{\beta}_L H }\mO e^{-\Bar{\beta}_R H}\right]$ is a normalization factor. These states can be prepared by the Euclidean path integral for a single CFT on a cylinder (Fig.~\ref{fig:contour}). We will focus on the case where $\mO$ is a product of $n$ scalar operators $\mO_\Delta$  uniformly supported on the $(d-1)$-dimensional spherical time slice of the CFT. In the dual gravitational theory, these operators $\mO$ create spherically symmetric shells of matter of mass $m=n m_\Delta$, where $n$ is taken to be of  $O\left(\frac{\ell^{d-1}}{G_N}\right)$ such that the shell is heavy enough to backreact classically on the geometry at leading order in $G_N$. 

Following \cite{Climent:2024trz}, we can project these states onto the microcanonical band $[E_{L,\, R},\, E_{L, \,R} + \delta E)$ by using a projector $\Pi_E = \Pi_E^L \otimes \Pi_E^R$,  
\be
| \Psi^E \rangle =\frac{1}{\sqrt{\langle \Psi|\Pi_E|\Psi\rangle}}\, \Pi_E | \Psi \rangle \,.\label{micro-state}
    \ee
We can now define the microcanonical bulk Hilbert space ${\cal H}^E_{\rm bulk}$ in terms of these normalized states. First, we consider a family of $\kappa$ states $|\Psi_i \rangle$, where operators $\mO_i$ with different conformal weights $\Delta_i$ create shells of different masses $m_i = n m_{\Delta_i}$. We project them into the microcanonical band as in (\ref{micro-state}) and use this family to introduce the auxiliary Hilbert space
\begin{equation}\label{eq:Kspanmicro}
    {\cal H}^E_{\rm bulk}(\kappa) \equiv {\rm Span}\{|\Psi^E_i\rangle, \quad i=1,\ldots,\kappa\}\,.
\end{equation}
As shown in \cite{Balasubramanian:2022gmo,Balasubramanian:2022lnw} and recapitulated below, when $\kappa$ is large enough, the vectors in this family become linearly dependent, and ${\cal H}^E_{\rm bulk}(\kappa) \to \mH^E_{\text{bulk}}$. Specifically, to leading order in the semiclassical approximation, the states $|\Psi^E_i \rangle$ and $|\Psi^E_j \rangle$ corresponding to  different operators $\mO_i$ and $\mO_j$ are orthogonal, i.e., $\overline{\langle \Psi^E_i|\Psi^E_j \rangle} \propto \delta_{ij}$, where the overbar denotes a coarse-grained gravitational saddlepoint calculation. However, wormhole contributions to the gravitational path integral induce a small overlap 
\cite{Balasubramanian:2022gmo,Balasubramanian:2022lnw,Climent:2024trz}. This non-orthogonality leads to saturation of the dimension of the linear span (\ref{eq:Kspanmicro}), defining the finite-dimensional microcanonical Hilbert space.  The dimension turns out to be precisely the exponential of the Bekenstein-Hawking entropy \cite{Balasubramanian:2022gmo,Balasubramanian:2022lnw,Climent:2024trz}. We will show that this mechanism is also responsible for factorization of the bulk Hilbert space into a product of left and right factors, generalizing \cite{Boruch:2024kvv}.

Here, we will work to leading order in \( G_N \), \( \kappa^{-1} \), and \( e^{-1/G_N} \). The $G_N$ is assumed to be small, ensuring that the semiclassical approximation to the gravitational path integral is valid. We will see that the appearance of the factorization property \eqref{eq: factorization-main} is connected to the convergence of the auxiliary Hilbert space ${\cal H}_{\rm bulk}^E (\kappa)$ to the true microcanonical Hilbert space ${\cal H}_{\rm bulk}^E$. This happens when  $\kappa \gtrsim e^{1/G_N}$~\cite{Balasubramanian:2022gmo}, so $\kappa$ is assumed to be large. We will specify more precisely the regimes required for different steps of the factorization argument below.

\subsection{Trace factorization and dimension of the microcanonical Hilbert space}\label{Trace factorization}

Following~\cite{Boruch:2024kvv},  factorization of the bulk Hilbert space in~\eqref{eq:Kspanmicro} can be studied by computing  the  trace ${\rm Tr}_{{\cal H}^E_{\rm bulk}}\left( k_L k_R\right)$. As discussed in  Appendix~\ref{sec:LinearAlgebra}, the microcanonical bulk trace can be defined in terms of our overcomplete basis of states spanning the bulk Hilbert space $\cal H_{\rm bulk}$ as\footnote{Note the difference in index contractions relative to~\cite{Boruch:2024kvv}. We use the expression derived in appendix~\ref{sec:LinearAlgebra}.}
\begin{equation}\label{eq:trace-K}
    {\rm Tr}_{{\cal H}^E_{\rm bulk}(\kappa)}\left( k_L k_R\right)= \left(G^{-1}\right)_{ji}\langle \Psi^E_i| k_L k_R |\Psi^E_j\rangle\,,
\end{equation}
where a sum on $i,j$ is implicit, and $G_{ij}\equiv \langle \Psi^E_i|\Psi^E_j\rangle$ is the Gram matrix of the family of states in~\eqref{eq:Kspanmicro}. The inverse Gram matrix $G^{-1}$ appears here to account for the non-orthogonality of the basis states.  The dimension of $G$ is $\kappa$, namely the number of states in the family \eqref{eq:Kspanmicro}, but the rank of $G$, namely the number of linearly independent states, can be smaller.  In the latter case $G$ has vanishing eigenvalues and hence is non-invertible, so $G^{-1}$ is understood as a generalized inverse where the non-zero eigenvalues are inverted, and the zero eigenvalues are left unchanged.\footnote{A priori, the generalized inverse of a matrix is unstable against tiny corrections if there are zero eigenvalues, i.e., if the matrix has less than maximal rank.   We are working in the leading semiclassical approximation, and hence we may worry that higher order corrections could change the rank of the matrix dramatically, by adding tiny corrections to many previously vanishing values. However, if this approach indeed provides an adequate account of the Bekenstein-Hawking entropy \cite{Balasubramanian:2022gmo,Balasubramanian:2022lnw}, we should expect that the effects of such corrections on the rank of the matrix should be small.  We leave it to future work to check this.} In effect this procedure projects out redundant states from the computation of the trace, while diagonalizing the remaining states that form a basis spanning the Hilbert space. It is  convenient to write the expression involving the inverse Gram matrix as an analytic continuation from positive powers of $G$ as follows\footnote{In this analytic continuation, we first observe that $0^n = 0$ for all $n>0$, and then continue the right hand side to $n=-1$, getting 0.}
\begin{equation}\label{eq:trace-K-bar}
     \overline{{\rm Tr}_{{\cal H}^E_{\rm bulk}(\kappa)}\left( k_L k_R\right)}=\lim_{n\to -1} \overline{\left(G^n\right)_{ji}\langle \Psi^E_i| k_L k_R |\Psi^E_j\rangle}\, .
\end{equation}
Here we also included an overbar as before to indicate that we will determine these quantities via a semiclassical saddle point computation.

The moments of the overlaps $\left(G^n\right)_{ij}$ can be computed by evaluating the resolvent
\begin{equation}\label{eq:resolve}
    R_{ij}(\lambda) = \left(\frac{1}{\lambda-G}\right)_{ij} = \frac{\delta_{ij}}\lambda + \frac{1}{\lambda} \sum_{n=1}^\infty \frac{\left(G^n\right)_{ij}}{\lambda^n}\,.
\end{equation}
The calculus of residues then tells us that 
\begin{equation}
    \label{eq:tracen}
    \overline{\left( G^n \right)_{ji}\langle \Psi^E_i| k_L k_R |\Psi^E_j\rangle} = \frac{1}{2\pi i} \oint d\lambda \lambda^n \overline{R_{ji}(\lambda)\langle \Psi^E_i| k_L k_R |\Psi^E_j\rangle}\,.
\end{equation}
We can calculate the resolvent and its trace, following \cite{Penington:2019kki,Balasubramanian:2022gmo,Balasubramanian:2022lnw,Climent:2024trz}, as reviewed below. In the semiclassical approximation, taking the trace of \eqref{eq:resolve} leads to
\begin{equation}\label{eq:resolve-bar}
    R(\lambda) = \frac{\kappa}{\lambda} + \frac{1}{\lambda} \sum_{n=1}^\infty  \sum_{i=1}^\kappa \frac{\overline{\left(G^n\right)_{ii}}}{\lambda^n}\,,
\end{equation}
where $R=\overline{{\rm tr}R_{ij}}$ and $\kappa$ is the dimension of the Gram matrix.  The moments of the overlap $\overline{\left(G^n\right)_{ii}}$ can be computed via the gravitational path integral and an inverse Laplace transform. This is a two-step procedure. First, we compute the overlap $\langle \Psi_i|\Psi_j\rangle$ of canonical states $|\Psi_{i,j}\rangle$~\eqref{state} via a path integral. In the semiclassical approximation, this overlap is given by the exponentiated on-shell action of classically allowed geometries filling the boundary conditions specified by the canonical states $|\Psi_{i,j}\rangle$. Next, we inverse Laplace transform the overlap to get the semiclassical approximation to the overlap of the microcanonical states $\overline{G_{ij}} = \overline{\langle \Psi^E_i|\Psi^E_j\rangle}$ since the canonical states are related to the microcanonical states $|\Psi_{i,j}^E\rangle$~\eqref{micro-state} by a Laplace transform. We refer the reader to Appendix \ref{app:states and gram} for more details.

The expansion in \eqref{eq:resolve-bar} can be represented diagrammatically as
\be
    \includegraphics[scale=0.9]{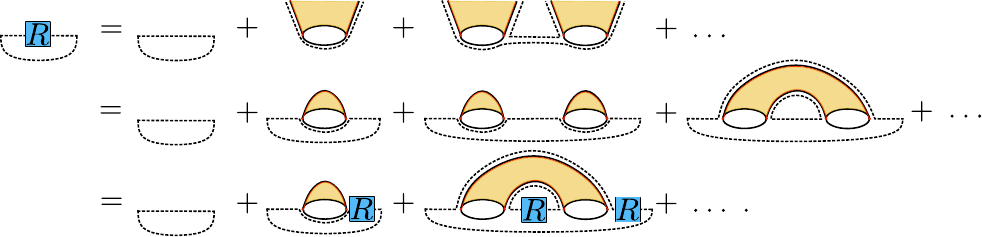} 
    \label{fig:SDR}
\ee
The diagrammatic expansion consists of inner products, shown by circles, chained together with dashed lines (representing the microstate indices) through the insertions of the internal shell that determines the microstate (shown by red dots). Upon applying the gravity path integral, the inner product circles set the boundary conditions for spacetime geometries to be filled in, as shown by shaded regions. Hence the diagrammatic expansion for the trace of the resolvent $R$ sums over geometries that fill in these boundary conditions. The resulting gravitational saddles are drawn in the second line. Note that when we are calculating a quantity that contains a product of more than one set of overlaps ($n>1$), there are several asymptotic boundaries, and the classically allowed geometries include connected and disconnected configurations. The third diagram in the second line is an example of a disconnected geometry contributing to a term with two overlaps, while the fourth diagram is a connected geometry that joins both asymptotic boundaries. We assume that $\kappa$ and $e^{1/G_N}$ are both large, without assuming any particular relationship between them. In this limit, we only need to consider planar geometries, as explained in~\cite{Penington:2019kki,Balasubramanian:2022gmo}. Non-planar geometries have higher topology and/or crossings, which are suppressed by powers of $e^{1/G_N}$ and $1/\kappa$, respectively.\footnote{ One might worry that because the solution to the Schwinger-Dyson equation~\eqref{eq:resolve-bar} is ${\cal O}(\kappa)$, non-planar contributions with large number of resolvent insertions could become important despite the $1/\kappa$ suppression. However, the planar diagrams in the third line of~\eqref{fig:SDR} have the largest number of index loops for the number of resolvent insertions. Non-planar diagrams with crossings are related to those by merging index loops, resulting in $1/\kappa$ suppressions while keeping the number of $R$ the same.} In the third line, the sum in equation~\eqref{eq:resolve-bar} is rearranged to contain only contributions from fully connected geometries at the cost of an $R$-insertion for each copy of the overlaps.
    
Thus the diagrammatic expansion leads to the Schwinger-Dyson equation
\begin{equation}\label{eq:resolve-bar-connected}
    R(\lambda) = \frac{\kappa}{\lambda} + \frac{1}{\lambda} \sum_{n=1}^\infty  \sum_{i=1}^\kappa  \frac{R^n}{\kappa^n} \overline{\left(G^n\right)_{ii}}\Big|_{\rm con}\,,
\end{equation}
where $\overline{\left(G^n\right)_{ii}}\Big|_{\rm con}$ is determined by the inverse Laplace transform of the gravitational on-shell partition function of the fully connected geometry that connects all the asymptotic boundaries, with the normalization ensuring that the states in the Gram matrix are normalized. For the microcanonical states in~\eqref{eq:Kspanmicro} in the heavy shell limit $m\to \infty$,\footnote{Note that \cite{Balasubramanian:2022gmo,Balasubramanian:2022lnw,Climent:2024trz} showed that a complete basis of microstates can be built from heavy internal shells of this kind.} the fully connected contribution is given by
\begin{equation}\label{eq:G-con}
    \sum_{i=1}^\kappa\overline{\left(G^n\right)_{ii}}\Big|_{\rm con} =\kappa^n e^{(1-n)(S_L+S_R)}\,,
\end{equation}
where $S_L=S(E_L)$ and $S_R=S(E_R)$.  As discussed in \cite{Balasubramanian:2022gmo,Balasubramanian:2022lnw} $S_{L,R}$ are equal to the microcanonical entropies counting the number of states in a single copy of the CFT in the respective energy window. For  details of the derivation of \eqref{eq:G-con}, see Appendix~\ref{tiny calculation} and \cite{Penington:2019kki,Balasubramanian:2022gmo,Balasubramanian:2022lnw,Climent:2024trz,Infalling}. Using \eqref{eq:G-con}, the Schwinger-Dyson equation~\eqref{eq:resolve-bar-connected} gives
\begin{equation}\label{eq:eqnresolvent}
     \lambda R= \kappa + \frac{R e^{S_L+S_R}}{e^{S_L+S_R} -R}\,.
\end{equation}
This quadratic equation in $R$ can be easily solved and generically has two solutions. The analytic properties of the resolvent can single out the correct solution, as we will explain below equation \eqref{eq:Trace}.

\begin{figure}[t]
    \centering
    \includegraphics[scale=1.4]{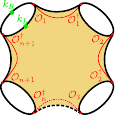} 
    \caption{The ($n+1$)-boundary wormhole geometry for large masses of the shells (red) with $k_L$ and $k_R$ inserted on a single asymptotic boundary (green).} 
    \label{fig:wormhole}
\end{figure}

We are now ready to proceed to the computation of the trace in~\eqref{eq:trace-K}. The integrand on the right-hand side of~\eqref{eq:tracen} can also be computed using the gravitational path integral via a Schwinger-Dyson equation. The resulting calculation is very similar to the one for the resolvent described above and pictorially takes the form 
\be
    \includegraphics[scale=1]{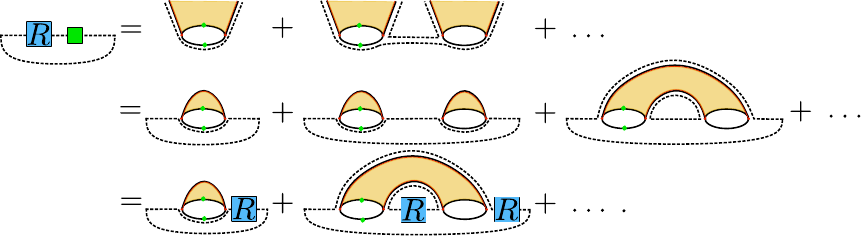}  
    \label{fig:SDtrace}
\ee
The new ingredient here is the green box, which denotes the insertion of $\langle \Psi_i | k_L k_R | \Psi_j \rangle$. On the right-hand side its effect translates to inserting the operators $k_L$ and $k_R$ on the respective asymptotic boundaries, as shown by the green dots. The third line in the diagrammatic equation~\eqref{fig:SDtrace} is a series expansion in powers of $R = \overline{\tr R_{ij}}$, where the coefficients can be computed using the gravity path integral and an inverse Laplace transform. The diagrammatic expansion can be written explicitly as 
\begin{equation}\label{eq:SD trace}
    \overline{R_{ji}(\lambda)\langle \Psi^E_i| k_L k_R |\Psi^E_j\rangle} = \sum_{n=0}^\infty R^{n+1} \frac{\bold{Z}_{n+1}(k_L,k_R)}{\bold{Z}_1^{n+1}}\,,
\end{equation}
where
\be\label{eq: inverse laplace of Z_n(k)}
\bZ_{n+1}(k_{L},k_R) = \sqrt{\frac{\mathbf{h}_{n+1}}{2\pi}} \int d\beta\  e^{(n+1)\beta E_{L,R}} Z_{n+1}(\beta; k_L, k_R)\,, 
\ee
and
\be Z_{n+1} (\beta; k_L, k_R) = e^{-I_{n+1}(\beta; k_L, k_R)}\,
\ee
in the saddlepoint approximation where $I_{n+1}(\beta; k_L, k_R)$ is the on-shell action of the $(n+1)$-boundary wormhole with $k_L$,$k_R$ insertions on one of the boundaries shown on Fig.~\ref{fig:wormhole}.\footnote{The $\bZ_n$ without $k$-insertions can be recovered by the relation 
\begin{equation}
    \bold{Z}_{n}(k_L=1,k_R=1) = \bold{Z}_{n}\,, 
\end{equation}
where $\bZ_n$ is defined by the inverse Laplace transform (\ref{eq: bZn}).} The quantity $\mathbf{h}_{n+1}$ in (\ref{eq: inverse laplace of Z_n(k)}) is the Hessian (second derivative) of the wormhole action $I_{n+1}$. The denominator in~\eqref{eq:SD trace} comes from the normalization of the $2(n+1)$ microcanonical states in each term of eq.~\eqref{fig:SDtrace}, as explained in appendix \ref{tiny calculation}.

The quantity $\bold{Z}_{n}$ with $k_L=k_R=1$ was extensively studied in~\cite{Balasubramanian:2022gmo,Balasubramanian:2022lnw,Climent:2024trz}. This calculation is readily generalized to arbitrary $k_{L,\,R}$ that are functions of the Hamiltonians. In the present section, we focus on the simple case where 
\begin{equation}\label{eq: choice klkr step0}
    k_L = e^{-\beta_L H_L}, \qquad k_R = e^{-\beta_R H_R}\,
\end{equation}
are Euclidean time evolution operators for arbitrary $\beta_L$ and $\beta_R$ on the left and right sides of the double-sided black hole. Then the necessary modification of the calculation of $\bZ_n$ amounts to increasing the length of the asymptotic boundaries on which the operators $k_L$ and $k_R$ are inserted by $\beta_L$ and $\beta_R$, respectively.  We present the details of this calculation in appendix~\ref{app:states and gram}. The trace (\ref{eq:trace-K-bar}), which we compute here via analytic continuation in the coarse-grained saddlepoint approximation, then becomes $\overline{\bZ(E, \beta)}$ defined in (\ref{eq: bZ-def}).  For the choice of operators \eqref{eq: choice klkr step0}, $\bZ_{n+1}(k_L, k_R)$ is given by 
\begin{equation}\label{eq:Zn+1}
    \frac{\bold{Z}_{n+1}(k_L\,,k_R)}{\bold{Z}_1^{n+1}}=e^{ -nS\left(E_L\right)-nS\left(E_R\right)-{\beta}_LE_L-{\beta}_RE_R}\,,
\end{equation}
where we recall that $S_L=S(E_L)$ and $S_R=S(E_R)$ are the microcanonical entropies that count the number of states in a single copy of the CFT in the respective energy window. The two main features of the result~\eqref{eq:Zn+1} that are important for our computation are that it factorizes between the left and right sides, and has simple $n$ dependence, which permits resummation of the Schwinger-Dyson equation~\eqref{eq:SD trace}. We refer the reader to appendix~\ref{tiny calculation} for details.

Plugging $\bZ_{n+1}(k_L,k_R)$ from~\eqref{eq:Zn+1} into the Schwinger-Dyson equation~\eqref{eq:SD trace} gives 
\begin{equation}\label{eq:trace-solved}
    \overline{R_{ji}(\lambda)\langle \Psi^E_i| k_L k_R |\Psi^E_j\rangle} = e^{-\beta_L E_L-\beta_R E_R} \frac{R(\lambda) e^{S_L+S_R}}{e^{S_L+S_R}-R(\lambda)}\,.
\end{equation}
We therefore find 
\begin{equation}
    \overline{\left( G^n \right)_{ji}\langle \Psi^E_i| k_L k_R |\Psi^E_j\rangle} =   \oint \frac{d\lambda}{2\pi i} \lambda^n e^{-\beta_L E_L-\beta_R E_R}  \frac{R(\lambda) e^{S_L+S_R}}{e^{S_L+S_R}-R(\lambda)}\,,
\end{equation}
such that
\begin{equation}\label{eq:Trace}
    \overline{{\rm Tr}_{{\cal H}^E_{\rm bulk}(\kappa)}\left( k_L k_R\right)} = \oint \frac{d\lambda}{2\pi i} \frac{1}{\lambda} e^{-\beta_L E_L-\beta_R E_R}  \frac{R(\lambda) e^{S_L+S_R}}{e^{S_L+S_R}-R(\lambda)}\,,
\end{equation}
where the contour over which we integrate encircles $\lambda=0$ clockwise as discussed in Fig.~1 of \cite{Boruch:2024kvv}. The details of this discussion rely on the analytic structure of the trace of the resolvent in the complex $\lambda$-plane, which we now consider.

From the definition of the resolvent~\eqref{eq:resolve}, it is clear that for a finite-sized matrix $G$, the resolvent contains poles at each eigenvalue of $G$, and the residue of each pole counts the degeneracy of the corresponding eigenvalue. Both solutions to~\eqref{eq:eqnresolvent} contain a branch cut, which can be thought of as an accumulation of poles in the continuum limit.\footnote{This continuum limit arises because there are $e^S$ poles along an interval of finite size, so their average separation is of order $e^{-S}$. In the semiclassical approximation, it is impossible to resolve the exact location of these poles to that precision, resulting in a branch cut approximating the accumulation of poles.} A crucial ingredient that singles out the correct solution is the degeneracy of the zero eigenvalue, which determines the residue of $R$ at $\lambda=0$. In equations, since the Gram matrix has $\kappa$ eigenvalues,\footnote{In this argument, we consider a degenerate eigenvalue with degeneracy $\ell$ to count as $\ell$ eigenvalues.} and they are all real and non-negative, this implies that
\begin{equation}\label{eq: R constraint}
    \oint \frac{d\lambda}{2\pi i} R(\lambda) = \kappa\,,
\end{equation}
where the contour is counter-clockwise and encircles the line $[0,\Lambda]$ where $\Lambda$ is large enough that all branch points and poles of the resolvent are in the region $|\lambda| < \Lambda$. Moreover, the trace of the Gram matrix gives $\kappa$, which also implies
\begin{equation}\label{second}
    \oint \frac{d\lambda}{2\pi i} \lambda R(\lambda)=\kappa,
\end{equation}
for the same contour as in~\eqref{eq: R constraint}.

Since the Schwinger-Dyson equation~\eqref{eq:eqnresolvent} is a simple quadratic equation for the trace of the resolvent, the solutions to this equation $R(\lambda)$ are straightforward to find. One can verify that only one solution satisfies \eqref{second} identifying the correct solution. We are particularly interested in the asymptotic behavior of this solution when $\lambda\to 0$, which is given by
\begin{equation}\label{eq:asymptoticsR}
R(\lambda)= \frac{\kappa-e^{S_L+S_R}}{\lambda}\Theta\left(\kappa-e^{S_L+S_R}\right)+R_0 + O(\lambda)\,,
\end{equation}
where $\Theta$ is the Heaviside step function and 
\begin{equation}\label{R0}
R_0=\frac{e^{S_L+S_R} \left(\left(e^{S_L+S_R}+\kappa \right) \Theta \left(\kappa -e^{S_L+S_R}\right)-\kappa \right)}{e^{S_L+S_R}-\kappa } \,.
\end{equation}
This is in agreement with~\cite{Balasubramanian:2022gmo,Boruch:2024kvv} and the following properties of this solution can be verified:
\begin{itemize}
    \item For $\kappa>e^{S_L+S_R}$, the degeneracy of the zero eigenvalue of the Gram matrix $G$ is $\kappa-e^{S_L+S_R}$. Since the residue of the resolvent at $\lambda=0$ corresponds to the number of states with zero eigenvalue, the solution satisfies $\text{ Res}_{\lambda=0}R(\lambda)=\kappa-e^{S_L+S_R}$.

    \item For $\kappa<e^{S_L+S_R}$, the Gram matrix has no zero eigenvalue and thus $R(\lambda)$ is regular at $\lambda=0$, specifically $R(0)=R_0$ with $R_0$ a finite number.  
\end{itemize}

With the expansion~\eqref{eq:asymptoticsR} for the resolvent, we now return to the trace in~\eqref{eq:Trace}. When $\kappa<e^{S_L+S_R}$, $R(\lambda)$ is regular at $\lambda=0$ and the integral in~\eqref{eq:Trace} simply picks up the residue, which leads to
\begin{equation}
     \overline{{\rm Tr}_{{\cal H}^E_{\rm bulk}(\kappa)}\left( k_L k_R\right)} =  e^{-\beta_L E_L-\beta_R E_R} \frac{R_0 e^{S_L+S_R}}{R_0-e^{S_L+S_R}}\,.
\end{equation}
While in the above equation, the $\beta_L$ and $\beta_R$ dependence factorizes, the $E_L$ and $E_R$ dependence does not (recall that $S_{L,R}=S(E_{L,R})$). Therefore the trace does not factorize into the product of a left and right trace as in \eqref{eq: bZ-def} for $\kappa < e^{S_L+S_R}$.

On the other hand, precisely when $\kappa>e^{S_L+S_R}$, the family of states in~\eqref{eq:Kspanmicro} becomes linearly dependent. In particular, the null states are responsible for a pole in the resolvent at $\lambda\to 0$, see \eqref{eq:asymptoticsR}. In this case, the residue at $\lambda=0$ in~\eqref{eq:Trace} is
\begin{equation}\label{eq: tracefactorizationstep0}
    \overline{{\rm Tr}_{{\cal H}^E_{\rm bulk}(\kappa)}\left( k_L k_R\right)}=\overline{\bZ(E; \beta)}  = e^{S_L+S_R- \beta_L E_L-\beta_R E_R}\,,
\end{equation}
where the first equality reflects the fact that for $\kappa > e^{S_L+S_R}$, the Hilbert space ${\cal H}_{\rm bulk}^E(\kappa)$ reproduces the true microcanonical Hilbert space ${\cal H}_{\rm bulk}^E$.  The expression on the right-hand side factorizes into a product of left and right pieces.  The factorization above was shown in semiclassical approximation of the low enegy theory.  In the next section we will show that it holds regardless of the details of the UV completion.

Note that using the AdS/CFT correspondence each piece in (\ref{eq: tracefactorizationstep0}) could be interpreted as a quantity in a single copy of a CFT, computed here using a semiclassical approximation in a dual gravity description. This can be done by first computing the thermal partition function of the CFT$_{L,R}$ with temperature $\beta+\beta_{L,R}$ using the gravitational path integral, and then inverse Laplace transform the temperature $\beta$ to energy $E_{L,R}$. This results in the microcanonical trace of $e^{-\beta_{L,R} E_{L,R}}$ over the respective Hilbert spaces\footnote{The right-hand side differs from the result in \cite{Boruch:2024kvv}. However, to leading order in $e^{-1/G_N}$, we have $\overline{{\rm Tr}_{{\cal H}_L^{E_L}}(k_L)}\, \overline{{\rm Tr}_{{\cal H}_R^{E_R}}(k_R)} = \overline{{\rm Tr}_{{\cal H}_L^{E_L}}(k_L) {\rm Tr}_{{\cal H}_R^{E_R}}(k_R)}$. We present the factorized form because we expect the factorization to hold beyond the leading semiclassical order.} 
\begin{equation}\label{eq:Zbar-factor2}
    \overline{\bZ(E; \beta)}  = \overline{{\rm Tr}_{{\cal H}_L^{E_L}}(k_L)}\ \overline{ {\rm Tr}_{{\cal H}_R^{E_R}}(k_R)}\,.
\end{equation}

This concludes our derivation of the main trace factorization result that will be central to the analysis below. Specifically, the above calculation demonstrates that the coarse-grained expression for $\overline{{\rm Tr}_{{\cal H}^E_{\rm bulk}(\kappa)}\left( k_L k_R\right)}$ factorizes to leading order in $e^{S_L+S_R}$ for the operators $k_{L,R}$ in \eqref{eq: choice klkr step0}, provided that $\kappa > e^{S_L+S_R}$. As discussed below~\eqref{eq: bZ-def}, the factorization of $\bZ(E; \beta)$ occurs if and only if the dimension of the microcanonical Hilbert space also factorizes. By comparing \eqref{eq: bZ-def} with \eqref{eq: tracefactorizationstep0}, we can directly extract the dimension of the Hilbert space and verify that it indeed factorizes into left and right components
\begin{equation}
    \overline{{\rm Tr}_{{\cal H}^E_{\rm bulk}(\kappa)}\left( \mathds{1}\right)}  = e^{S_L+S_R}\,.
\end{equation}

\subsection{Trace differential and fine-grained factorization}\label{Trace differential}

Having proven that the coarse-grained expression for $\overline{{\rm Tr}_{{\cal H}^E_{\rm bulk}}\left( k_L k_R\right)}$ factorizes to leading order in $e^{-1/G_N}$, we now turn our attention to showing factorization at the fine-grained level. To this end, we will use the differential matrix
\begin{equation}\label{eq: diff}
    d(E; \beta)_{\alpha_L\,\alpha_R} = \bZ (E; \beta) \partial_{\alpha_L} \partial_{\alpha_R} \bZ (E; \beta)-\partial_{\alpha_L} \bZ (E; \beta) \partial_{\alpha_R} \bZ(E; \beta) \,,
\end{equation}
where $\alpha_{L,R} \in \{\beta_{L,R},E_{L,R}\}$. As discussed in Sec.~\ref{sec: recap} and footnote~\ref{foot:dE=0}, the differential matrix vanishes if and only if the trace factorizes into left and right traces at the fine-grained level. The $\beta_L \beta_R$ component of the differential
\begin{equation}
\begin{aligned}
   d(E; \beta)_{\beta_L\,\beta_R} & = {\rm Tr}_{{\cal H}^E_{\rm bulk}} \left(k_L k_R \right) {\rm Tr}_{{\cal H}^E_{\rm bulk}}\left(H_L k_L k_R H_R\right) - {\rm Tr}_{{\cal H}^E_{\rm bulk}}\left(H_L k_L k_R \right)  {\rm Tr}_{{\cal H}^E_{\rm bulk}}\left(k_L k_R H_R \right)
\end{aligned}
\end{equation}
trivially vanishes because the Hamiltonians evaluate to $E_{L,R}$, respectively, in the microcanonical bands and can be taken out of the traces.\footnote{As a side note, we can show that
\begin{equation}
\begin{aligned}
   &{\rm Tr}_{{\cal H}^E_{\rm bulk}}\left(H_L k_L k_R H_R\right) - \frac{{\rm Tr}_{{\cal H}^E_{\rm bulk}}\left(H_L k_L k_R \right)  {\rm Tr}_{{\cal H}^E_{\rm bulk}}\left(k_L k_R H_R  \right)}{{\rm Tr}_{{\cal H}^E_{\rm bulk}} \left(k_L k_R\right) }\\
   &= \langle E | H_L k_L k_R  H_R | E \rangle  - \frac{ \langle E | H_L k_L k_R  | E \rangle \langle E |  k_L k_R H_R| E \rangle }{\langle E|k_L k_R|E\rangle}\,
\end{aligned}
\end{equation}
where $|E\rangle$ is the microcanonical thermofield double~\cite{Marolf:2018ldl}, see eq.~\eqref{eq: TFDmicro}. This state is equal to the microcanonical states in eq.~\eqref{micro-state} with the operator ${\cal O}$ set to the identity; see appendix~\ref{app:states and gram} for more details. Hence, the fact that the $\beta_L\beta_R$ component of the differential matrix vanishes is equivalent to the vanishing of the connected two-point function of the Hamiltonian in the state $\sqrt {k_L k_R}|E\rangle$.} To show that the other components of the differential matrix vanish, we exploit the fact that if their second moments $\overline{(d(E;\beta_L)_{\alpha_L\, \alpha_R})^2}$  vanish, then the differential matrix vanishes at the fine-grained level to the orders in $e^{-1/G_N}$ and $G_N$ to which our computation is valid. 

To compute the differential matrix, we write 
\begin{equation}
    \overline{d(E;\beta)_{\alpha_L\,\alpha_R}} = \mathcal{D}_{\alpha^{1,2}_{L/R}}\overline{ \bZ(E^1; \beta^1) \bZ(E^2;\beta^2)}\bigg|_{\alpha^{1,2}_{L,\,R}=\alpha_{L,\,R}}\,,
\end{equation}
\begin{equation}
    \overline{\left(d(E;\beta)_{\alpha_L\,\alpha_R}\right)^2} = \mathcal{D'}_{\alpha^{1,2,3,4}_{L/R}}\overline{ \bZ(E^1; \beta^1)\bZ(E^2; \beta^2)\bZ(E^3; \beta^3)\bZ(E^4; \beta^4)}\bigg|_{\alpha^{1,2,3,4}_{L,\,R}=\alpha_{L,\,R}}\,,
\end{equation}
where 
\begin{equation}
   \mathcal{D}_{\alpha^{1,2}_{L/R}}=\partial_{\alpha^1_L}\partial_{\alpha^1_R}-\partial_{\alpha^1_L}\partial_{\alpha^2_R}\,,
\end{equation}
and 
\begin{equation}
    \mathcal{D'}_{\alpha^{1,2,3,4}_{L/R}}=\partial_{\alpha^1_L}\partial_{\alpha^1_R}\partial_{\alpha^2_L}\partial_{\alpha^2_R}-2\partial_{\alpha^1_L}\partial_{\alpha^1_R}\partial_{\alpha^2_L}\partial_{\alpha^3_R}+\partial_{\alpha^1_L}\partial_{\alpha^2_R}\partial_{\alpha^3_L}\partial_{\alpha^4_R}\,.
\end{equation}

Therefore, we need to compute the average of multiple traces  $\overline{ \bZ(E^1; \beta^1)\ldots \bZ(E^n; \beta^n)}$. As usual, this can be done using the gravitational path integral. For $n=2$, this quantity can be computed by summing the diagrams

\be
    \includegraphics[scale=0.9]{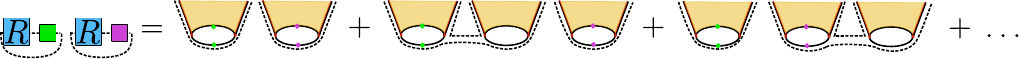} 
    \label{fig:doubletrace}
\ee
Similarly to figure~\ref{fig:SDtrace}, the green box and purple box denote the insertion of $\langle \Psi_i | k^1_L k^1_R | \Psi_j \rangle$ and $\langle \Psi_i | k^2_L k^2_R | \Psi_j \rangle$, respectively. Recall also that each term on the right hand side specifies boundary conditions to fill in with classically allowed geometries.  Importantly, for each of these boundary conditions, there are contributions that connect the boundary conditions associated with different copies of the two-sided partition function by a wormhole. For example, in the planar limit, the second term above receives contributions from the following geometries 
\be
    \includegraphics[scale=1]{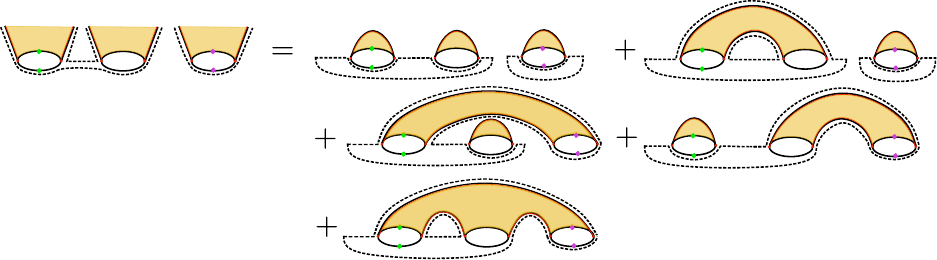} 
    \label{fig:doubletrace2}
\ee
The contributions connecting the two copies of the two-sided partition function, depicted in the last two lines of the diagrams, represent geometries that would, in principle, spoil the factorization of the product of partition functions. However, in the limit where $\kappa$ becomes very large, these contributions are suppressed by powers of $\kappa^{-1}$, rendering them subleading compared to the dominant diagrams. For this example, the leading contribution comes from the second diagram on the right-hand side, where the traces remain disconnected. This suppression of connected geometries can be confirmed by counting the number of index loops in each diagram, with each loop contributing a factor of $\kappa$. As a result, the diagrams that contribute to leading order in $\kappa^{-1}$ are those where the two traces are entirely independent.

Hence, in what follows we work in the planar limit and then take $\kappa \to \infty$, which significantly reduces the number of diagrams that need to be considered. This is in contrast to the previous section, where $\kappa$ and $e^{1/G_N}$ were treated as large but uncorrelated parameters. Here, we consider $\kappa$ to be much larger than $e^{1/G_N}$, allowing us to span the full Hilbert space once $\kappa>e^{S_L+S_R}$. After this point, further increasing 
$\kappa$ merely adds null states, leaving the physical Hilbert space unchanged. It is important to note that $\kappa$ is not a physical parameter but a computational tool measuring the dimension of the auxiliary matrix $G$, with its rank corresponding to the Hilbert space dimension. This approach simplifies the analysis without altering any conclusions. Further implications of the planar limit and the role of $\kappa$ will be discussed in Sec.~\ref{sec: discuss}.

With this setup, we can now demonstrate the factorization of the coarse-grained trace product in the planar limit as $\kappa\to\infty$. Following the same procedure used to compute the coarse-grained trace in Sec.~\ref{Trace factorization}, we arrive at the factorized form for two copies of the partition function
\begin{equation}
    \overline{ \bZ(E^1; \beta^1)\bZ(E^2; \beta^2)}=\overline{ \bZ(E^1; \beta^1)}\quad \overline{\bZ(E^2; \beta^2)}
    \,,
\end{equation}
Because this result was derived using the planar limit, it is valid to leading order in $e^{-1/G_N}$.\footnote{In deriving this factorization, we used the limit $\kappa\to \infty$ to suppress the contributions from many subleading diagrams. However, once $\kappa$ is large enough to span the entire microcanonical Hilbert space, the trace becomes independent of $\kappa$. Thus, for finite $\kappa$ larger than $e^{S_L+S_R}$,  the sum of these subleading diagrams must vanish for every power of $\kappa^{-1}$ to reproduce the correct answer. Proving this in general is a more intricate question, but it is not needed to show the Hilbert space factorization to leading order in $e^{-1/G_N}$.} The generalization to arbitrary $n$ can be argued analogously and leads to 
\begin{equation}
    \overline{ \bZ(E^1;\beta^1)\ldots \bZ(E^n; \beta^n)}=\overline{ \bZ(E^1;\beta^1)}\,\ldots\, \overline{ \bZ(E^n; \beta^n)}
    \,, \label{eq: factorization-multiboundary}
\end{equation}
which, once again, is valid to leading order in $e^{-1/G_N}$.

We can now apply the factorization of each individual microcanonical two-sided partition function~\eqref{eq: tracefactorizationstep0}, proven in Sec.~\ref{Trace factorization}, to (\ref{eq: factorization-multiboundary}) and get 
\begin{equation}\label{eq: factorization}
    \overline{ \bZ(E^1; \beta^1)\ldots \bZ(E^n; \beta^n)}= e^{S^1_L+S^1_R- \beta^1_L E^1_L-\beta^1_R E^1_R}\,\ldots \, e^{S^n_L+S^n_R- \beta^n_L E^n_L-\beta^n_R E^n_R}\,,
\end{equation}
where $S^i_{L,R} = S(E^i_{L,R})$. Considering the case $n=4$, the differential operator $\mathcal{D'}_{\alpha^{1,2,3,4}_{L/R}}$ acting on (\ref{eq: factorization}) gives zero. Thus, we have shown to leading order in $e^{-1/G_N}$ and $\kappa^{-1}$, that the second moment of each component of the differential matrix vanishes $\overline{(d(E;\beta)_{\alpha_L\,\alpha_R})^2}=0$. As explained in Sec.~\ref{sec: recap}, this implies that the trace factorizes into a left and a right part as in eq.~\eqref{eq: important}. Following the proof outlined in Sec.~\ref{sec:proof}, one concludes from this that the bulk microcanonical Hilbert space factorizes, which in turn implies the factorization of the Hilbert space and 
\begin{equation}
    Z_{\rm bulk}=\Tr_{\mH_L}  k_L\Tr_{\mH_R} k_R \, .
\end{equation}
This is our main result. Note that this result is independent of holography and solely relies on the gravitational path integral to show that the bulk Hilbert space factorizes as a product $\mH_{\rm bulk}=\mH_L \otimes\mH_R$.

Moreover, we can leverage the result in \eqref{eq: factorization-multiboundary} to also show that the variance of $\bZ$ vanishes,
\begin{equation}
    \overline{\left(\bZ(E;\beta)-\overline{\bZ(E;\beta)}\right)^2}=0\,,
\end{equation}
to leading order in $e^{-1/G_N}$. This implies that to leading order in $e^{-1/G_N}$, without averaging over UV degrees of freedom, the microcanonical trace
\begin{equation}\label{eq: Z=Zbar}
    \bZ(E;\beta) = \overline{\bZ(E;\beta)}\,.
\end{equation}
The averaged microcanonical trace $\overline{\bZ}$ includes tracing over UV degrees of freedom to reduce to the low energy effective description of GR before taking the trace in the microcanonical Hilbert space of this EFT. Without the averaging, $\bZ$ lacks this trace over UV degrees of freedom and instead depends on the exact configuration of these degrees of freedom while the IR degrees of freedom are still traced over. The statement $\bZ=\overline{\bZ} + {\cal O}(e^{-1/G_N})$ specifies that the value of the IR trace is not very sensitive to the UV degrees of freedom. 

\section{Trace factorization implies Hilbert space factorization}
\label{sec:proof}
In this section, we show how trace factorization implies Hilbert space factorization, under certain conditions.  See also the comments in the Discussion section of \cite{Boruch:2024kvv}.

Let $H_1$ and $H_2$ be two commuting Hermitian operators on a Hilbert space $\cal H$, i.e.,
\begin{equation}
    [H_1,H_2]=0.
\end{equation}
This implies that the operators can be simultaneously diagonalized so that there exists a common orthonormal basis  $\{|\psi_{\lambda_a \mu_b}\rangle\}$ for ${\cal H}$,
\begin{equation}
    H_1|\psi_{\lambda_a \mu_b}\rangle=\lambda_a|\psi_{\lambda_a \mu_b}\rangle\,,\qquad H_2|\psi_{\lambda_a \mu_b}\rangle=\mu_b|\psi_{\lambda_a \mu_b}\rangle\,.
\end{equation}
In our context, $H_{1,2}$ will be the left and right bulk Hamiltonians which are dual to the boundary CFT Hamiltonians. 
We will first assume that no other operators commute with $H_1$ and $H_2$ and hence that the eigenspace corresponding to each pair of eigenvalues $(\lambda_a, \mu_b)$ is non-degenerate, and will later explain how the reasoning can be extended to relax this assumption. Under this assumption, the eigenspaces of $H_1$ are only degenerate because of the presence of $H_2$, and vice versa.

Next, suppose that the trace of any functions of these operators factorizes as
\begin{equation}\label{factorizationassumption}
    {\rm Tr}_{\cal H} \left(f(H_1) g(H_2)\right)=\frac{ {\rm Tr}_{\cal H} f(H_1) \, {\rm Tr}_{\cal H} g(H_2)}{ {\rm Tr}_{\cal H} \mathds{1}}.
\end{equation}
Now we can use the fact that for each Hermitian operator, there exist operators that project onto the eigenspaces corresponding to each of their eigenvalues.  Specifically, for a finite dimensional Hilbert space we can write
\begin{equation}
P_a({H_1}) = \prod_{\lambda \neq \lambda_a} \frac{H_1 - \lambda \mathds{1}}{\lambda_a - \lambda}\,,\qquad P_b({H_2}) = \prod_{\mu \neq \mu_b} \frac{H_2 - \mu \mathds{1}}{\mu_b - \mu} \, .
\end{equation}
In terms of these projection operators, we can write
\begin{equation}
    H_1=\sum_a^n \lambda_i P_a(H_1)\,,\qquad H_2=\sum_b^m \mu_b P_b(H_2)\,,
\end{equation}
where $n$ and $m$ label the number of distinct eigenvalues of $H_1$ and $H_2$, respectively.

Since each projection operator can be expressed as a function of the respective operator $H_1$ and $H_2$, we can use the factorization assumption in \eqref{factorizationassumption} to deduce that
\begin{equation}\label{eq: trace projector}
    {\rm Tr}_{\cal H}\left( P_a(H_1) P_b(H_2)\right)=\frac{{\rm Tr}_{\cal H} P_a(H_1)   {\rm Tr}_{\cal H} P_b(H_2) }{ {\rm Tr}_{\cal H} \mathds{1}}=\frac{d_{\lambda_a}d_{\mu_b}}{d}\,,
\end{equation}
where $d$ is the dimension of $\cal H$ and $d_{\lambda_a}$ is the degeneracy of the eigenvalue $\lambda_a$ of $H_1$ and similarly for $d_{\mu_b}$.
Equation~\eqref{eq: trace projector} implies that for each pair of eigenvalues $(\lambda_a\,,\mu_b)$, the corresponding eigenspace is at least one-dimensional, ${\rm Tr}_{\cal H}\left( P_a(H_1) P_b(H_2)\right)>0$.
Now, recall that we have assumed the absence of additional symmetries, i.e., no operator other than $H_1$ and $H_2$ commute with both $H_1$ and $H_2$. This condition ensures that there are no additional degeneracies in the eigenvalue spectrum of the operators, ${\rm Tr}_{\cal H}\left( P_a(H_1) P_b(H_2)\right)<2$. As a result, for each pair of eigenvalues $(\lambda_a\,,\mu_b)$, the corresponding eigenspace is exactly one-dimensional,
\begin{equation}
    {\rm Tr}_{\cal H} P_a(H_1) P_b(H_2)=1 \quad \forall \lambda_a\,,\mu_b\,.
\end{equation}

This in turn implies that 
\begin{equation}\label{eq: Hspan}
    {\cal H} =  {\rm Span} \left\{ |\psi_{\lambda_a \mu_b} \rangle\  |\  \lambda_a \in {\rm eigenval}(H_1)\,, \mu_b \in {\rm eigenval}(H_2) \right\}\,.
\end{equation}
The non-trivial consequence of trace factorizing is that the list of eigenvalues in \eqref{eq: Hspan} includes states with all possible combinations of $(\lambda_a,\mu_b)$, which is a necessary requirement for the Hilbert space to be a tensor product. Compatibility of the inner product follows from the orthogonality of states with different eigenvalues of $H_1$ and/or $H_2$. It is also easy to argue that any linear operator on the Hilbert space $\mH$ necessarily has a representation as sums of products of operators acting separately on the two factors.  This follows from counting the dimensions of the operator spaces on the Hilbert space and its factors, at least for finite dimensional Hilbert spaces, such as the microcanonical black hole Hilbert spaces.

Let us now go back to the problem at hand. In Sec.~\ref{Trace differential} we have shown that for the bulk microcanonical Hilbert space, the trace will factorize \eqref{eq: important}. Hence, the argument presented above implies that the bulk microcanonical Hilbert space will factorize into a direct product of two smaller Hilbert spaces when no additional conserved quantities are present. To extend the argument beyond the microcanonical ensemble, consider larger left and right energy bands. We want to show that for each combination of left and right energy eigenvalues, the bulk Hilbert space contains a single corresponding eigenstate. To do so, observe that we can simply choose left and right microcanonical energy windows centered around the chosen eigenvalues and apply the above argument again.  

In conclusion, we have shown that the entire Hilbert space $\mH$ factorizes into a direct product of smaller Hilbert spaces 
\begin{equation}
    \mH=\mH_1\otimes\mH_2\,,
\end{equation}
where $\mH_1$ is spanned by the eigenstates of $H_1$ and $\mH_2$ by the eigenstates of $H_2$ when no additional conserved quantities are present.  

Let us now comment on the assumption of no additional charges that commute with $H_{1,2}$, which we took to be left and right bulk Hamiltonians.  In quantum gravity, we indeed do not expect any conserved charges associated with global symmetries.  However, there can be conserved charges associated with gauge symmetries. To derive trace factorization in the presence of such charges, we need to find that there are states in the Hilbert space associated with any choice of left or right charge and that these are not constrained by some relation arising from the gauge symmetry \cite{Harlow_2016,Guica_2017,Boruch:2024kvv}.  This can be shown using the generalized charged shell microstates constructed in \cite{Climent:2024trz}.  These are guaranteed to exist as long as there are charge sources associated with the gauge field, which can be arranged to produce charged shells propagating behind the horizon.  For any bulk theory with a CFT dual, we are guaranteed that operators creating such states exist.\footnote{In AdS/CFT a bulk gauge symmetry is dual a boundary global symmetry.  If the CFT does not contain operators charged under this symmetry, then it acts trivially and is not a symmetry at all.}

This concludes our argument that the Hilbert space of the two-sided black hole in AdS space factorizes at least to leading non-perturbative order, and in any spacetime dimension for theories with a consistent Euclidean gravity path integral, and general relativity as the low energy limit.

\section{Discussion}\label{sec: discuss}

In this paper, we use the gravitational path integral  to show factorization of the bulk Hilbert space of the eternal black hole to leading order in the semiclassical approximation. Specifically, we show  that the trace $\Tr_{\mH_{\text{bulk}}} k_L k_R$, where $k_{L,R}$ are dual to functions of the boundary Hamiltonians,  factorizes to leading order in the expansion in $e^{-1/G_N}$.  We then show that this trace factorization implies Hilbert space factorization. The argument is based on inserting a vastly overcomplete family of semiclassically well-defined microstates into the trace. The factorized form of the bulk trace then arises as a consequence of the fact that the dimension of the bulk Hilbert space spanned by these microstates is given by the entropy of the two-sided black hole. While Hilbert space factorization followed from trace factorization of a restricted family of operators, in Appendix~\ref{sec: generic} we argue that the trace factorization argument also applies directly to the trace of  general light operators with limited backreaction on the saddle point geometries of the gravity path integral. 

Our results are valid 
to leading order in the expansion in $e^{-1/G_N}$ of the gravitational path integral of GR with asymptotically AdS$_{d+1}$ boundary conditions for $d \geq 2$. This formal gravity path integral should be thought of as an effective low-energy description of a UV-complete theory of gravity. 
By contrast, the authors of \cite{Boruch:2024kvv} are able to work to all orders in $e^{-1/G_N}$ because they are treating two-dimensional JT gravity, which is exactly solvable, including its nonperturbative corrections \cite{Saad:2019lba}.  In general dimensions, the nonperturbative structure is not fully understood, which makes it difficult to classify subleading contributions.

A  key step in our work is the microcanonical projection: we first restrict the bulk trace computation  to  microcanonical bands and then combine the results for different bands to demonstrate our result. In JT gravity \cite{Boruch:2024kvv} showed trace factorization for finite-width energy bands directly by including the Laplace transform between canonical and microcanonical pictures as part of the computation of state overlaps. 

We compute our results in the
$\kappa \to \infty$ limit, where $\kappa$ is the number of semiclassical microstates (in a given microcanonical band) included in the family (\ref{eq:Kspanmicro}). In \cite{Balasubramanian:2022gmo} it was shown that as long as $\kappa \gtrsim e^{1/G_N}$, this family of states is an overcomplete basis for the Hilbert space, and increasing $\kappa$ beyond the value corresponding to the black hole entropy only adds null states without affecting any physical observables. It is thus a virtue of the resolvent computations discussed in Sec. \ref{Trace factorization} and \ref{Trace differential} that making $\kappa$ large eliminates most of the diagrams except for the planar ones, simplifying the calculation drastically. Moreover, because the physical quantities are independent of $\kappa$ after the appearance of null states, this implies that, while the sum of the diagrams at each order of $\kappa^{-1}$ contributes to the Schwinger-Dyson equation for the resolvent~\eqref{eq:resolve-bar-connected}, it does not affect the final answer for both the resolvent~\eqref{eq:asymptoticsR} and the trace~\eqref{eq: tracefactorizationstep0}. Thus, while the $\kappa \to \infty$ limit is an important mathematical trick to simplify the computation, the physical result remains general.

We can also consider the relation between our approach and the factorization puzzle of Euclidean gravity observables \cite{Maldacena:2004rf,Saad212,Saad2021,Cotler2021,Blommaert2021, Belin2021,Penington:2023dql,Hernandez-Cuenca2024} on a set of disconnected boundaries.\footnote{This factorization puzzle is in contrast to the question addressed in this work, which concerns the factorization of the operator algebra of two distinct boundaries in the Lorentzian description. When continued to the Euclidean geometry, the corresponding operators are all located on one Euclidean asymptotic boundary, as shown in figure~\ref{fig:wormhole}. The factorization puzzle we discuss in this paragraph concerns multiple Euclidean asymptotic boundaries.} A common manifestation of this puzzle is the apparent tension between the Euclidean connected spacetime multi-boundary wormhole contributions, which dominate at the coarse-grained level, and  factorization of the product of multiple two-sided partition functions of the form (\ref{eq: factorization-multiboundary}) \cite{Maldacena:2004rf}. In Sec.~\ref{Trace differential} we demonstrated the factorization of the coarse-grained product of two-sided partition functions to leading order in the semiclassical approximation. This was needed to complete the proof of the bulk Hilbert space factorization to leading order in $e^{-1/G_N}$ and $\kappa^{-1}$, similarily to the JT gravity case \cite{Boruch:2024kvv}. Thus in the present work we have shown that, to leading order in $e^{-1/G_N}$, both the bulk Hilbert space factorization and the factorization of Euclidean observables on disconnected boundaries hold. This is in contrast to the work in JT gravity~\cite{Boruch:2024kvv}, which addressed the Hilbert space factorization problem to all orders, but the Euclidean gravity factorization puzzle only to leading order.

Finally, note that the overcomplete family of semiclassical microstates introduced in \cite{Balasubramanian:2022gmo}, which has been the key tool in our argument, can also be constructed in asymptotically flat spacetimes and for one-sided black holes \cite{Balasubramanian:2022lnw} (see also \cite{Geng:2024jmm}). This suggests that our results can be generalized to eternal black holes in flat space. It would also be interesting to understand if such semiclassically well-controlled microstates can be constructed in  de Sitter space, potentially providing an interpretation to the entropy associated to the cosmological horizon and perhaps providing a mechanism for factorization between static patches of de Sitter.

\section*{Acknowledgements}

We would like to thank Will Chan, Shira Chapman, Charlie Cummings, Chitraang Murdia, Martin Sasieta, Alejandro Vilar L\'opez, Gabriel Wong and Tom Yildrim for useful discussions. VB was supported in part by the DOE through DE-SC0013528 and QuantISED grant DE-SC0020360, and in part by the Eastman Professorship at Balliol College, Oxford. Work at VUB was supported by FWO-Vlaanderen project G012222N and by the VUB Research Council through the Strategic Research Program High-Energy Physics. JH is supported by FWO-Vlaanderen through a Junior Postdoctoral Fellowship. 

\appendix

\section{Traces in non-orthogonal and overcomplete bases}
\label{sec:LinearAlgebra} 

\paragraph{Gram matrix: } Suppose we have a $d$ dimensional Hilbert space $\mathcal{H}$ spanned by $n \geq d$, generally non-orthogonal, basis elements $ \mathcal{V} = \{ |v_i\rangle \}$.  Consider the Gram matrix 
\begin{equation}
G_{ij} = \langle v_i | v_j \rangle\,.
\end{equation}
$G$ is Hermitian, and so can be diagonalized by a unitary matrix $U$,
\begin{equation}
G = U D U^{\dagger} \, ,
\label{eq:diagonalGram}
\end{equation}
where $D$ is a diagonal matrix with real, positive semidefinite eigenvalues $\lambda_i$. 
 The rank of $G$ equals the number of positive eigenvalues, which is the dimension of the Hilbert space $d$.  The difference between the dimension and rank of $G$, $n-d$, equals the number of zero eigenvalues, and measures the dimension of the null space of $G$.  Powers of the Gram matrix,
\begin{equation}
G^p = U D^p U^{\dagger} \, ,
\end{equation}
are well defined for any real $p \geq 0$, where $D^p = {\rm diag}(\lambda_1^p, \lambda_2^p,\ldots \lambda_d^p, 0,0,\ldots 0)$.   We will understand $G^p$ for negative $p$ by continuation from positive $p$.  For example, analytically continuing to $p=-1$ gives $G^{-1} = U D^{-1} U^\dagger$ with $D^{-1} = {\rm diag}(\lambda_1^{-1}, \lambda_2^{-1},\ldots \lambda_d^{-1}, 0,0,\ldots 0)$.  So $\Tr(G^{-1} G) = \sum_{jk} G^{-1}_{kj}G_{jk} = \Tr(U D^{-1} D U^{\dagger})$.  Using the cyclic property of the trace we get $\Tr(G^{-1} G) = \Tr(D^{-1} D)$.  Now because $D^{-1}$ is understood by analytic continuation as described above, $\Tr \left(D^{-1} D\right) =  \Tr({\rm diag}(1,\ldots 1, 0,\ldots))$, where the first $d$ diagonal entries are $1$ and the rest are $0$. So 
\begin{equation}
\Tr(G^{-1} G) =  \Tr(D^{-1} D ) = d \, .
\end{equation}
In effect the null space has been projected out.  Similarly, $G^{-{1 \over 2}} = U D^{-{1 \over 2}} U^\dagger$ with $D^{-{1 \over 2}} = {\rm diag}(\lambda_1^{-{1 \over 2}}, \lambda_2^{-{1 \over 2}},\ldots, \lambda_d^{-{1 \over 2}}, 0,0,\ldots, 0)$.  Since $G$ is Hermitian so is $G^p$, so that $G^{p\dagger} = G^p$.

\paragraph{Orthonormal basis: } An orthonormal basis for the Hilbert space can be constructed via the linear combinations
\begin{equation}
|u_a\rangle = \sum_{ij}  G^{-{1\over2}}_{ji} U_{ia} |v_j\rangle  = \sum_{j b}  U_{jb}D_{ba}^{- {1\over 2}} |v_j\rangle\,,\qquad \langle u_a | = \sum_{ij} \langle v_j | G_{ji}^{-{1\over 2} *} U_{ia}^* =
\sum_{bj} \langle v_j | D_{ab}^{-{1\over 2} } U^{\dagger}_{bj}\, ,
\end{equation}
where in the last equation we used the Hermiticity of $G$.  Indeed, we can check that the first $d$ $|u_a\rangle$ are orthonormal, and the remaining ones are simply zero
\begin{equation}
\langle u_a | u_b \rangle =
\sum_{jkcd}D_{ac}^{-{1\over 2}} U^{\dagger}_{cj} \langle v_j| v_k \rangle U_{kd} D_{db}^{-{1\over 2}}
=
\sum_{cd}D_{ac}^{-{1\over 2}} D_{cd} D_{db}^{-{1\over 2}} = \left(\mathds{1}_d\right)_{ab} \, ,
\end{equation}
where $\mathds{1}_d = {\rm diag}(1,1,\ldots,1,0,0,\ldots,0)$ projects out the null states in the last $n-d$ entries. This implies that we are free to project out the null states and let $a,b,\ldots$ indices run from $1$ to $d$.

\paragraph{Inner products: } We can use the orthonormal basis $|u_a\rangle$ to derive a formula for inner products in the original non-orthogonal basis $\mathcal{V}$.  For any states $|A\rangle$ and $|B\rangle$ we can write
\begin{equation}
\langle A | B \rangle =
\sum_a \langle A |u_a\rangle \langle u_a| B \rangle =
\sum_{abcjk} \langle A  |v_j\rangle U_{jb} D^{-{1\over 2}}_{ba} D^{-{1\over 2}}_{ac} U^{\dagger}_{ck}\langle v_k| b \rangle
= \sum_{jk}   
\langle A |v_j\rangle G^{-1}_{jk} \langle v_k| B \rangle \, .
\end{equation}

\paragraph{Traces: } We can similarly derive a formula for the trace of an operator $\mathcal{O}$:
\begin{equation}\label{eq:traceformula}
\Tr{\mathcal{O}} = 
\sum_a \langle u_a | \mathcal{O} | u_a \rangle =
\sum_{abcjk} D^{-{1\over 2}}_{ab} U^{\dagger}_{bj} \langle v_j| {\cal O}| v_k \rangle U_{kc} D^{-{1\over 2}}_{ca} =
\sum_{jk} G^{-1}_{kj}  \langle v_j | \mathcal{O}  
 | v_k \rangle \, .
\end{equation}
As a check, note that if $\mathcal{O}=I$ then $\Tr(\mathcal{O}) = d$. Noting that $\langle v_j | I| v_k \rangle = I_{jk} = G_{jk}$ we see that (\ref{eq:traceformula}) gives $\sum_{jk} G^{-1}_{kj} I_{jk} = \sum_{jk} G^{-1}_{kj} G_{jk} = d$. 

\section{Microstates and overlaps}\label{app:states and gram}

This appendix explains the relation between canonical and microcanonical versions of the family of microstates introduced in Sec.~\ref{Basis of states}, and discusses their overlaps.

\subsection{Canonical and microcanonical thermofield double}

In this section we look at a simple example, the thermofield double. We explain the connection between the microcanonical thermofield double and the canonical thermofield double using a Laplace transform and compute their various overlaps. 

The canonical thermofield double states at temperature $\beta$ are given by
\begin{equation}
    |\beta \rangle = \frac{1}{\sqrt{Z(\beta)}}\sum_n e^{-\beta E_n/2} |n \rangle \otimes |n \rangle\,.
\end{equation}
These states can be prepared by a Euclidean path integral, and their normalization constant is the thermal partition function
\begin{equation}
    Z(\beta) ={\rm Tr}\ \e^{-\beta H}\,.
\end{equation}
The overlap between two thermofield doubles is
\begin{equation}\label{eq:b-overlap}
    \langle \beta | \beta' \rangle = \frac{Z\left((\beta+\beta')/2\right)}{\sqrt{Z(\beta) Z(\beta')}}\,.
\end{equation}

To define the microcanonical thermofield double states, one projects the canonical states into a microcanonical energy band using a projector $\Pi_E$:
\begin{equation}
    |E_\beta\rangle = \sqrt{Z(\beta)e^{\beta E-S(E)}} \Pi_E | \beta \rangle = e^{-S(E)/2}\sum_{E_n \approx E} |n,n\rangle\,, \label{eq: TFDmicro}
\end{equation}
where $\sqrt{Z(\beta)e^{\beta E-S(E)}}$ normalizes the state $|E_\beta\rangle $ after the projection. Additionally, 
\begin{equation}
    e^{S(E)} = \sum_{E_n\approx E} 1
\end{equation} 
is the dimension of the microcanonical Hilbert space in the band of energy $[E,E+\delta E)$ with $\delta E \ll \beta^{-1}$, and correspondingly $S(E)$ is the microcanonical entropy. 
The microcanonical thermofield double does not depend on the temperature of the thermofield double that was projected into the energy band. Hence, we omit the $\beta$ label from now on. The above state is similar to the $\sigma \to 0$ limit of the microcanonical thermofield double state considered in~\cite{Marolf:2018ldl}, which allows for energy windows centered around $E$ and supported by some function $f(M)$, which is chosen to be a Gaussian $f(M) = \exp\left( -(M-E)^2/4\sigma^2\right)$.

The overlap between the microcanonical thermofield double and the canonical thermofield double is
\begin{equation}\label{eq:bE-overlap}
    \langle \beta | E \rangle = \frac{e^{S(E)/2-\beta E/2}}{\sqrt{Z(\beta)}}\,.
\end{equation}
The overlap between two microcanonical thermofield doubles is
\begin{equation}\label{eq:E-overlap}
    \langle E | E' \rangle = \delta_{E E'}\,.
\end{equation}

The microcanonical thermofield double can be obtained from the thermofield double by an inverse Laplace transform\footnote{Note that if we want to match the above result to the state in~\cite{Marolf:2018ldl}, there is a factor of $-i$ missing. This is just a phase which doesn't affect the state itself, so we omit it here.
}
\begin{equation}
   | E \rangle=\sqrt{\frac{\mathbf{h}_{1/2}}{2\pi }}\int d\beta e^{\beta E/2-S(E)/2} \sqrt{Z(\beta)} |\beta \rangle\,,
\end{equation}
where the integration is performed along the complex contour going from $-i\infty$ to $+i\infty$ remaining to the right of any singularities. The Hessian $\mathbf{h}_{1/2}$ corresponds to the second derivative of the on-shell action $I_{1/2}=\frac{1}{2}\log Z$, evaluated on the saddle point value of $\beta$. 

The overlaps can also be obtained by inverse Laplace transforms:
\begin{equation}
    \langle \beta | E \rangle = \sqrt{\frac{\mathbf{h}_{1/2}}{2\pi }} \int d\beta' e^{\beta' E/2-S(E)/2} \sqrt{Z(\beta')} \langle \beta |\beta' \rangle\,,
\end{equation}
which agrees with~\eqref{eq:bE-overlap} by using~\eqref{eq:b-overlap}. The microcanonical overlaps can also be computed using an inverse Laplace transform
\begin{equation}
    \langle E | E'\rangle = \frac{\mathbf{h}_{1/2}}{2\pi } \int d\beta d\beta'e^{\frac{\beta E-S(E)+\beta'E'-S(E')}{2}}\sqrt{Z(\beta) Z(\beta')}\langle \beta|\beta'\rangle\,,
\end{equation}
which reproduces~\eqref{eq:E-overlap}.
Lastly, note that the norm of the microcanonical thermofield double can be computed by an inverse Laplace transform of the norm of the canonical thermofield double
\begin{equation}
    \langle E | E\rangle = \sqrt{\frac{\mathbf{h}_{1}}{2\pi }}\int d\beta e^{\beta E-S(E)} Z(\beta) \langle \beta | \beta \rangle = 1\,,
\end{equation}
where $\mathbf{h}_1$ is the Hessian of the on-shell action $I_1=\log Z$ evaluated at the saddle point value of $\beta$.

\subsection{Microstates with heavy shells}

Having introduced the microcanonical thermofield double as a warm-up exercise, we now move to the canonical states used in~\cite{Balasubramanian:2022gmo,Balasubramanian:2022lnw,Climent:2024trz}. These can be prepared by a path integral similar to the thermofield double, with the addition of an insertion of a heavy spherically symmetric operator $\mO$ which sources a non-interacting shell of matter in the bulk description. They are higher dimensional generalizations of the partially entangled thermal states (PETS) defined in JT gravity~\cite{Goel:2018ubv}.

In particular, the canonical states introduced in~\cite{Balasubramanian:2022gmo,Balasubramanian:2022lnw,Climent:2024trz} are
\begin{equation}\label{eq: shell state}
    |\Psi \rangle = \frac{1}{\sqrt{Z_1}}\sum_{n\,m} e^{-\frac{1}{2} \bar{\beta}_L E_m - \frac{1}{2} \bar{\beta}_R E_n} \mO_{mn} |m\rangle 
 \otimes |n\rangle\,,
\end{equation}
where $\mO_{mn} = \langle m | O |n\rangle$, and $Z_1=\text{Tr}\left[ \mO^\dagger e^{- \Bar{\beta}_L H }\mO e^{-\Bar{\beta}_R H}\right]$ normalizes these states. The operator $\mO$ is spherically symmetric with conformal dimension $\Delta \gg 1$. Because of the insertion of the heavy operator $\mO$, the expression now depends on two individual Euclidean time evolutions $\bar{\beta}_L$ and $\bar{\beta}_R$. The overlap of two such states with the same operator $\mO$ but different Euclidean time evolutions is (in the large $\Delta$ limit)
\begin{equation}\label{eq:canon-overlap}
   \langle \Psi | \Psi'\rangle = \frac{Z((\bar{\beta}_L+\bar{\beta}_L')/2) Z((\bar{\beta}_R+\bar{\beta}_R')/2)}{\sqrt{Z(\bar{\beta}_L)Z(\bar{\beta}_R)Z(\bar{\beta}_L')Z(\bar{\beta}_R')}} \,.
\end{equation}

The microcanonical version of these states can be found by projecting onto the left and right energy bands $[E_{L,R}, E_{L,R}+\delta E)$ with $\delta E \ll \bar{\beta}_{L,R}^{-1}$, which gives
\begin{equation}
\begin{aligned}
    |\Psi_E \rangle & =  \sqrt{\frac{Z_1 e^{\bar{\beta} E}}{{\rm Tr}_{{\cal H}^E}(\mO^\dagger \mO)}} \Pi_{E_L,E_R}|\Psi \rangle = \frac{1}{\sqrt{{\rm Tr}_{{\cal H}^E}(\mO^\dagger \mO)}} \sum_{E_m\approx E_L,E_n\approx E_R} \mO_{mn} |m\rangle\otimes|n\rangle\,,
\end{aligned}
\end{equation}
where we have introduced the shorthand notation $\bar{\beta} E=\bar{\beta}_L E_L+\bar{\beta}_R E_R$. The normalization constant is
\begin{equation}
    {\rm Tr}_{{\cal H}^E}(\mO^\dagger \mO) = \sum_{E_m \approx E_L, E_n \approx E_R} \mO^\dagger_{mn} \mO_{mn}\,.
\end{equation}
This normalization constant, which we denote as $\bZ_1$, is related to the normalization of the canonical states by an inverse Laplace transform
\begin{equation}
    \bZ_1 = {\rm Tr}_{{\cal H}^E}(\mO^\dagger \mO) = \frac{\mathbf{h}_{1}}{2\pi } \int d^2\bar{\beta} \, e^{\bar{\beta}_L E_L + \bar{\beta}_R E_R} Z_1\,. \label{eq: bZ1}
\end{equation}
Here we have used the notation $d^2\bar{\beta} = d\bar{\beta}_L d\bar{\beta}_R$.

The microcanonical state is related to the canonical state by an inverse Laplace transform 
\begin{equation}
    |\Psi^E\rangle = \frac{1}{\sqrt{\bold{Z}_1}} \frac{\mathbf{h}_{1/2}}{2\pi } \int d^2\bar{\beta} e^{\bar{\beta} E/2} \sqrt{Z_1} |\Psi \rangle\,.
\end{equation}
The overlap between canonical and microcanonical states is
\begin{equation}
    \langle \Psi | \Psi^E\rangle = \sqrt{\frac{\bold{Z}_1}{Z_1 e^{\bar{\beta}E}}} \,,
\end{equation}
and 
the overlap between microcanonical states is\label{eq: microcanon-overlap}
\begin{equation}
    \langle \Psi^E | \Psi^{E'} \rangle =\delta_{E_L E'_L}\,.
\end{equation}
Similarly to the thermofield double case, the microcanonical overlaps can be obtained from the canonical overlaps by an inverse Laplace transform. For example, one can find that the norm is given by
\begin{equation}
    \langle \Psi^E|\Psi^E \rangle = \frac{1}{\bold{Z}_1} \frac{\mathbf{h}_{1}}{2\pi } \int d^2\bar{\beta} e^{\bar{\beta}_L E_L+\bar{\beta}_R E_R} Z_1\langle \Psi | \Psi \rangle\,.
\end{equation}

\section{Family of states and overlaps from the path integral}
\label{tiny calculation}

In this appendix, we consider the family of microcanonical states used to define the bulk Hilbert space. 
We define the Gram matrix, and compute its moments in the semiclassical approximation.  We also explain how the insertion of the operators $k_{L,R}$ modifies the result. 

To this end, we turn our attention to canonical states, whose overlaps can be computed using the gravitational path integral. Each overlap adds an asymptotic boundary as part of the boundary conditions. The path integral is then approximated by its saddle point values, which consist of classically allowed solutions with those boundary conditions,
\begin{equation}\label{eq: semiclassical overlaps}
      \overline{\langle \Psi_{k_1} | \Psi_{k_2} \rangle \langle \Psi_{k_2} | \Psi_{k_3} \rangle  \ldots \langle \Psi_{k_n} | \Psi_{k_1}\rangle} \approx \frac{\sum_{\rm n- bdy\ saddles} e^{-I}}{\left(\sum_{\rm 1- bdy\ saddles} e^{-I}\right)^n}\,.
\end{equation}
The numerator in~\eqref{eq: semiclassical overlaps} computes the overlap structure using the gravity path integral, and the denominator is the normalization factor keeping the states $| \Psi_i \rangle $ normalized.

In this prescription, for $n=1$ we have by construction
\begin{equation}
    \overline{\langle \Psi_i | \Psi_j\rangle} = \delta_{ij} \, .
\end{equation}
For $n>1$, several asymptotic boundaries set the boundary conditions for the gravity path integral. The leading contributions in the semiclassical approximation from the connected saddles are nontrivial,
\begin{equation}
    \overline{\langle \Psi_i | \Psi_j \rangle \langle \Psi_j | \Psi_i\rangle} - \overline{\langle \Psi_i | \Psi_j \rangle} \ \overline{\langle \Psi_j | \Psi_i\rangle} = \frac{Z(2 \bar{\beta}_L) Z(2\bar{\beta}_R)}{Z( \bar{\beta}_L)^2 Z(\bar{\beta}_R)^2}\,.
\end{equation}
We label this expression for the connected contribution to the overlap squared as $\overline{\langle \Psi_i | \Psi_j \rangle \langle \Psi_j | \Psi_i\rangle}\big|_{\rm con}$. Generalizing to the $n$-th moment of the overlap, we write the leading connected contribution as 
\begin{equation}
   \overline{\langle \Psi_{k_1} | \Psi_{k_2} \rangle \langle \Psi_{k_2} | \Psi_{k_3} \rangle  \ldots \langle \Psi_{k_n} | \Psi_{k_1}\rangle}\big|_{\rm con} = \frac{Z_n}{Z_1^n}\,,
\end{equation}
where $Z_n$ is determined by the on-shell action of the $n$-boundary wormhole geometry. For the microstates in the heavy shell mass limit, the ratio reads
\be
\frac{Z_n }{Z_1}= \frac{Z(n\bar{\beta}_L) Z(n\bar{\beta_R})}{Z(\bar{\beta}_L) Z(\bar{\beta}_R)}\,. \label{eq: Zn-pinching}
\ee 
It is  straightforward  to inverse Laplace transform the overlaps of the canonical states to find the overlaps of the microcanonically projected states. When all states in the overlap are projected to the same microcanonical band $E$, the connected contribution to the $n$-th moment of the overlap has the form
\begin{equation}
   \overline{\langle \Psi^E_{k_1} | \Psi^E_{k_2} \rangle \langle \Psi^E_{k_2} | \Psi^E_{k_3} \rangle  \ldots \langle \Psi^E_{k_n} | \Psi^E_{k_1}\rangle}\big|_{\rm con} = \frac{\bZ_n}{\bZ_1^n}\,, \label{eq: n-th moment}
\end{equation}
where 
\begin{equation}
    \bZ_n = \frac{\mathbf{h}_{n}}{2\pi } \int d^2\bar{\beta} \, e^{\bar{\beta}_L E_L + \bar{\beta}_R E_R} Z_n\,, \label{eq: bZn}
\end{equation}
with $\mathbf{h}_n$ being the Hessian of the on-shell action for the $n$-boundary wormhole, and $\bZ_1$ is given by (\ref{eq: bZ1}). In the heavy shell mass limit, this leads to
\be
\frac{\bZ_n}{\bZ_1^n} = e^{(1-n)S(E_L)+(1-n)S(E_R)}\,. \label{eq: bZ-frac}
\ee
Thus, the semiclassical approximation to the connected part of the $n$-th power of the Gram matrix is
\begin{equation}
\overline{\left(G^n\right)_{i j}}\Big|_{\rm con} = \sum_{k_i}\overline{G_{i k_1}G_{k_1 k_2}\ldots G_{k_{n-1} j}}\Big|_{\rm con} = \kappa^{n-1}\delta_{ij} e^{(1-n)S(E_L)+(1-n)S(E_R)}\,,
\end{equation}
and in particular, its trace is 
\begin{equation}
\sum_{i=1}^{\kappa}\overline{\left(G^n\right)_{ii }}\Big|_{\rm con} =  \kappa^n \frac{\bold{Z}_{n}}{\bold{Z}_1^n} = \kappa^n e^{(1-n)S(E_L)+(1-n)S(E_R)}\,,
\end{equation}
where we used (\ref{eq: bZ-frac}) in the last step for the heavy shell mass limit. 

Finally, we can also evaluate the quantities appearing in the Schwinger-Dyson equation for the trace using
\begin{equation}
    \overline{\langle \Psi_{k_1}^E |k_L k_R | \Psi^E_{k_2}\rangle \langle \Psi^E_{k_2} | \Psi^E_{k_3} \rangle  \ldots \langle \Psi^E_{k_{n+1}} | \Psi_{k_1}\rangle}\big|_{\rm con} =   \frac{\bold{Z}_{n+1}(k_L\,,k_R)}{\bold{Z}_1^{n+1}}\,,
\end{equation}
where
\begin{equation}
    \frac{\bold{Z}_{n+1}(k_L\,,k_R)}{\bold{Z}_1^{n+1}}=e^{ -nS\left(E_L\right)-nS\left(E_R\right)-{\beta}_LE_L-{\beta}_RE_R}\,,
\end{equation}
which is used in \eqref{eq:Zn+1}.

\section{Factorization of the trace of generic operators}\label{sec: generic}

\begin{figure}[t]
    \centering
    \includegraphics[scale=1.4]{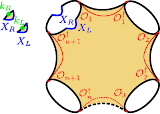} \hspace{2cm}
    \includegraphics[scale=1.4]{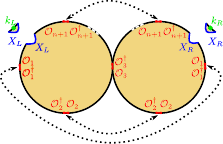} \\
    (a) \hspace{5cm} (b)
    \caption{(a) The $n+1$ boundary cutout wormhole geometry with operators $k_L$ and $k_R$ inserted on a single asymptotic boundary (green). The red curves are the shell worldlines. The operator insertions are cut out using codimension-one surfaces $X_L$ and $X_R$ (blue). (b) Same geometry in the limit of large masses of the shells. The dashed arrows show the gluing along the now pinched shell worldlines.}
    \label{cutout wormhole}
\end{figure}

To show the tensor product structure of Hilbert space we only needed to show trace factorization of the functions of conserved operators. Nevertheless, it is interesting to directly analyze trace factorization for arbitrary opertors $k_L$ and $k_R$, not necessarily functions of conserved charges. This is challenging in general because generic matter operators can backreact  on the geometry,  may interact with each other, and can lead to saddle points with different topologies.  Nevertheless, we can readily analyze operators that have limited backreaction on the saddle point geometries, similarly to \cite{Boruch:2024kvv}.  

Specifically, we focus on operators $k_{L, R}$ that are light enough that their backreaction does not alter the overall structure of the on-shell solutions. In particular, for light enough operators, the saddle point geometries to the gravity path integral are similar to the solutions without the insertion of the operators, and are accompanied by a field configuration determined by propagating the boundary conditions into the geometry with the appropriate Green's function. Importantly, the bulk field configuration backreacts on the saddle point geometries, giving solutions that are slightly different than the vacuum geometries. However, because the Green's functions decay away from the insertion point of an operator, the backreaction of the bulk field configuration on the on-shell geometry eventually becomes negligible far enough from the operator insertion. We will delineate the location at which this backreaction becomes negligible by some surfaces $X_{L,R}$, and will split the geometry into a small region that depends on the operators $k_{L,R}$, and the remaining region where the geometry matches the vacuum solution, which we refer to as the cutout wormhole, see figure~\ref{cutout wormhole}(a). The action then consists of an operator-dependent contribution ${\cal I}(k_{L,R};X_{L,R})$, as well as the universal contribution coming from the cutout wormhole $\mathbf{I}_{n+1}(E_L,E_R;X_L,X_R)$:
\begin{equation}\label{assumption step1}
    \bold{Z}_{n+1}(k_L,k_R) = e^{-\mI_{n+1}(k_L,E_L;X_L)} e^{-\mI_{n+1}(k_R,E_R;X_R)}  e^{- \bold{I}_{n+1}(E_L,E_R;X_L,X_R)}\,.
\end{equation}
The separation of the action  into operator-dependent terms and the universal contribution is enough to enable a factorization mechanism similar to the main text, if we add some assumptions about the cutout surfaces $X_{L,R}$ and the different contributions to the on-shell action as follows.
\begin{itemize}
    \item[(a)] We assume that the regions within the cutout surfaces, deformed by the operator backreaction within the vicinity of the corresponding insertions, are small and located close to the asymptotic boundary. The cutout surfaces $X_{L,R}$ never intersect any of the shell worldvolumes. This ensures that the contributions of the backreacted geometry to the on-shell action are universal: they depend only on the operators $k_{L,R}$ and bounding surfaces $X_{L,R}$, but not on the number of boundaries $n$,
    \begin{equation}
        {\cal I}_{n+1}(k_{L,R},E_{L,R};X_{L,R}) =  {\cal I}(k_{L,R},E_{L,R};X_{L,R})\,.
    \end{equation}
    The geometry of the wormhole away from the cutouts is the same as without the operator insertions: it consists of segments of the Euclidean AdS$_{d+1}$ spacetime, glued together along the shell worldvolumes. 
    
    \item[(b)] The heavy shell insertions are assumed to be much heavier than the operators $k_{L,R}$. The masses of the shells are taken to be heavy enough so that the shell worldvolumes can be pinched along the red shell worldvolumes, and the on-shell action of the geometry can be separated into contributions from the (deformed) Euclidean AdS black hole segments \cite{Balasubramanian:2022gmo}. In Fig.~\ref{cutout wormhole}(b), this pinching limit separates the front side and the back side of the cutout wormhole into two Euclidean disks. One of these disks has the $n+1$ replicated copies of CFT$_L$ on its asymptotic boundary, while the other one has the $n+1$ replicated copies of CFT$_R$. Therefore in the pinching limit the action of the cutout wormhole can be written as a sum of contributions depending on left and right parameters:
    \begin{align}\label{eq: assumtionstep1crusial}
    \lim_{m\to \infty} \bold{I}_{n+1}(E_L,E_R;X_L,X_R) &= \bold{I}_{n+1}(E_L;X_L) + \bold{I}_{n+1}(E_R;X_R)\,. 
    \end{align}
    \item[(c)] Assumptions (a) and (b) imply that the universal part of the action $\bold{I}_{n+1}$ is independent of $n$:
    \begin{equation}
    \bold{I}_{n+1}(E_{L,R};X_{L,R}) =-s_{L,R}(E_{L,R};X_{L,R})\,. \label{eq: ILR}
    \end{equation}
    To argue for $n$ independence, we begin with the case when $k_{L,R}(H_{L,R})$ are functions of the left and right Hamiltonians respectively. By comparing with the results of Sec.~\ref{sec:microcanonical}, we see\footnote{By comparing with the work of~\cite{Climent:2024trz}, we can conclude that a similar statement holds for the cases when $k_{L,R}$ are functions of extensive conserved quantities, specifically electric charge and angular momentum.}
    \begin{equation}
        s_{L,R}(E_{L,R};X_{L,R})-{\cal I}(k_{L,R},E_{L,R};X_{L,R})=S_{L,R}+ \log(k_{L,R}(E_{L,R}))\,.
    \end{equation}
   Specifically, for the case $k_{L,R}=\mathds{1}$, this implies
    \begin{equation}
        s_{L,R}(E_{L,R};X_{L,R}) = S_{L,R} + {\cal I}(k_{L,R}=\mathds{1},E_{L,R};X_{L,R})\,,
    \end{equation}
where ${\cal I}(k_{L,R}=\mathds{1},E_{L,R};X_{L,R})$ are the contributions of the on-shell action on the cutout regions for the case where the operators $k_{L,R}$ are trivial. Clearly this expression for $s_{L,R}(E_{L,R};X_{L,R})$ does not depend on $n$. But since $s_{L,R}(E_{L,R};X_{L,R})$ does not depend on the operator, it is in general true that the universal contribution $s_{L,R}$ is independent of $n$.
\end{itemize}

In order to enable the bulk trace computation as before, we can now write the general form of the $\bZ_{n+1}(k_L,k_R)$ for the cutout $(n+1)$-boundary wormhole together with the operator-dependent segments for general operators, such that the above assumptions are satisfied:
\begin{equation}\label{assumption step2}
    \bold{Z}_{n+1}(k_L,k_R) = e^{-\mI(k_L;X_L)} e^{-\mI(k_R;X_R)}e^{s_L(E_L;X_L)+s_R(E_R;X_R)}\,
\end{equation}
at leading order in $e^{-1/G_N}$. 

With the general form of $\bold{Z}_{n+1}(k_L,k_R)$ (\ref{assumption step2}), we can now compute $\overline{\bold{Z}(E;k)}$ using the same procedure as outlined in Sec.~\ref{Trace factorization}. To this end, we first insert~\eqref{assumption step2} into the Schwinger-Dyson equation for the trace~\eqref{eq:SD trace}, which yields\footnote{Note that this is the step where assumption (a) of $\mI$ not depending on $n$ is crucial}
\begin{equation}
    \overline{R_{ji}(\lambda)\langle \Psi^E_i| k_L k_R |\Psi^E_j\rangle} =  e^{-\mI(k_L;X_L)} e^{-\mI(k_R;X_R)} \frac{ R(\lambda) e^{s_L+s_R}}{e^{S_L+S_R}-R(\lambda)}\,.
\end{equation}
Next, we use \eqref{eq:trace-K-bar} and  \eqref{eq:tracen}, which gives
\begin{equation}\label{eq:Trace2}
    \overline{{\rm Tr}_{{\cal H}^E_{\rm bulk}(\kappa)}\left( k_L k_R\right)}  =  e^{-\mI(k_L,E_L;X_L)} e^{-\mI(k_R,E_R;X_R)}  \oint \frac{d\lambda}{2\pi i} \frac{1}{\lambda} \frac{ R(\lambda) e^{s_L+s_R}}{e^{S_L+S_R}-R(\lambda)}\,.
\end{equation}

The final ingredient to compute $\overline{\bold{Z}(E;k)}$ is the analytic structure of the trace of the resolvent in the complex $\lambda$-plane given in \eqref{eq:asymptoticsR}. If $\kappa<e^{S_L+S_R}$, $R(\lambda)$ is regular at $\lambda=0$ and the integral in~\eqref{eq:Trace2} picks up the residue
\begin{equation}
     \overline{{\rm Tr}_{{\cal H}^E_{\rm bulk}(\kappa)}\left( k_L k_R\right)} = e^{-\mI(k_L,E_L;X_L)} e^{-\mI(k_R,E_R;X_R)}  \frac{ R_0 e^{s_L+s_R}}{e^{S_L+S_R}-R_0}\,,
\end{equation}
where $R_0$ is defined in \eqref{R0}. This expression does not factorize.

On the other hand, when $\kappa>e^{S_L+S_R}$, the residue at $\lambda=0$ in~\eqref{eq:Trace2} is
\begin{equation}\label{eq: tracefactorization step1}
    \overline{{\rm Tr}_{{\cal H}^E_{\rm bulk}(\kappa)}\left( k_L k_R\right)} =\overline{\bold{Z}(E;k)}  = e^{-\mI(k_L,E_L;X_L)} e^{-\mI(k_R,E_R;X_R)}  e^{s_L+s_R}\,,
\end{equation}
which takes the desired factorized form. Specifically, the above calculation shows that the coarse-grained expression of $\overline{\bold{Z}(E;k)}$ factorizes to leading order in $e^{-1/G_N}$ for operators that satisfy \eqref{assumption step2} and \eqref{eq: assumtionstep1crusial}. To extend this discussion beyond the coarse-grained level, one can consider the family of operators
\be
k_{L,R}^a \in \{e^{a_{L,R} Q_{L,R}}|\, a_{L,R}\in \mathds{R}\}\,,
\ee
where $Q_{L,R}$ is a pair of fixed operators, and $a_{L,R}$ are real parameters enumerating the operator family such that $k_{L,R}^1=k_{L,R}$. Then one can use this operator family to define a differential matrix similar to \eqref{eq: diff-intro}, where now $\alpha_{L,R}\in \{E_{L,R},a_{L,R}\}$.   Using the same arguments as in Sec.~\ref{Trace differential}, one can then show that $\bZ(E;k)$ factorizes at the fine-grained level.

\bibliographystyle{JHEP}
\bibliography{factorization.bib}
\end{document}